\documentclass[5p,times]{elsarticle}
\usepackage[USenglish]{babel}
\makeatletter
\def\ps@pprintTitle{%
	\let\@oddhead\@empty
	\let\@evenhead\@empty
	\def\@oddfoot{\centerline{\thepage}}%
	\let\@evenfoot\@oddfoot}
\makeatother
\usepackage{lineno,hyperref}
\usepackage[utf8]{inputenc}   
\usepackage{epstopdf}           
\usepackage{physics}
\usepackage{flushend}  
\usepackage{amsfonts}
\usepackage{amsmath}
\usepackage{float}
\usepackage[dvipsnames]{xcolor}
\usepackage{graphicx}
\usepackage{color}
\usepackage{amssymb}
\usepackage{esint}
\usepackage{empheq}
\usepackage{calrsfs}
\usepackage{ragged2e}
\usepackage{subcaption}
\usepackage{kantlipsum} 
\usepackage{mwe}
\graphicspath{ {./Figures/} }
\biboptions{sort&compress}

\DeclareCaptionLabelSeparator{periodspace}{.\quad}
\modulolinenumbers[5]

\begin{document}
	
	\begin{frontmatter}
		
	\title{\vspace{-1.5cm}Soliton dynamics of a high-density Bose-Einstein condensate \\ subject to a time varying anharmonic trap \tnoteref{t1} }
	\tnotetext[t1]{© 2020. This manuscript version is made available under the CC-BY-NC-ND 4.0 license \url{http://creativecommons.org/licenses/by-nc-nd/4.0/}}
	\author[Ciencias,ICN]{R. Flores-Calderón \corref{cor1}}
	\ead{rflorescalderon@ciencias-unam.mx rafael.flores@correo.nucleares.unam.mx}
	
	\cortext[cor1]{Corresponding author}
	
	\author[IF]{J. Fujioka }
	\ead{ fujioka@fisica.unam.mx}
	
	\author[Ciencias]{A. Espinosa-Cerón }
	
	\address[Ciencias]{Facultad de Ciencias,Universidad Nacional Autónoma de México, Ciudad de México,CP 04510, México}
	\address[ICN]{
		Instituto de Ciencias Nucleares, Universidad Nacional Autónoma de México, Ciudad de México, CP 04510, México}
	\address[IF]{Departamento de Sistemas Complejos, Instituto de Física, Universidad Nacional Autónoma de México, Ciudad de México, CP 04510, México}
	
	\begin{abstract}
	In this paper we study the soliton dynamics of a high-density Bose-Einstein condensate (BEC) subject to a time-oscillating trap. The behavior of the BEC is described with a modified Gross-Pitaevskii equation (mGPE) which takes into account three-body losses, atomic feeding and quantum fluctuations (up to a novel high-density term). A variational approximation (VA) is used to study the behavior of a Gaussian pulse in a static double-well potential. Direct numerical solutions of the mGPE corroborate that the center of the pulse exhibits an oscillatory behavior (as the VA predicts), and show a novel phenomenon of fragmentation and regeneration (FR). It is shown that this FR process is destroyed if we consider a potential with a time-dependent quadratic term, but the FR survives if the time dependence is introduced in a cubic term. Comparison between the VA and the numerical solution revealed an excellent agreement when the oscillations of the pulse remain in one of the potential wells. The effects of the quantum fluctuating terms on the FR process are studied. Finally, variational results using a supergaussian trial function are obtained. 
	\end{abstract}
	
	\begin{keyword}
		\texttt{Dissipative solitons\sep Bose-Einstein Condensate \sep Variational Method \sep Three-body recombination \sep External feeding \sep Quantum fluctuations}
	\end{keyword}
	
\end{frontmatter}


\section{Introduction}

Recent progress on atom chips and quantum computers has turned the attention to strongly compressed quantum systems \cite{PhysRevLett-compress-quantum}. Candidates for these new technologies include topological materials and atomic gas systems such as Bose-Einstein condensates (BECs). For this reason, the study of the formation and propagation of matter waves results of utter importance. Therefore, in this paper we investigate the soliton dynamics and stability of high-density condensates. An important topic in the study of Bose-Einstein condensates (BECs) is the behavior of solitary waves in effectively one dimensional (1D) BECs with attractive interactions between atoms. Such waves can appear in narrow cylindrical BECs, confined by strong transverse and weak longitudinal magnetic traps. Solitary BEC's waves placed in such traps may exhibit harmonic oscillations if the longitudinal trapping potential has a parabolic profile. And a wider range of possibilities opens up if time-dependent potentials are considered. In these cases, the behavior of BECs may be more complex, and new interesting effects may appear. If we consider low-density BECs, their behavior in these traps can be satisfactorily described by using the Gross-Pitaevskii (GP) equation. On the other hand, in order to describe higher-density BECs it is necessary to modify the GP equation.

In the context of low temperatures, the Gross-Pitaevskii equation (GPE) has been shown to provide an effective approximation for the mean-field dynamics of the BEC wavefunction in many situations \cite{Ablowitz}. In a general condensate in which three-body losses, atomic feeding and high density is present, it needs to be modified to have  theoretically accurate predictions. The GPE assumes low density and no losses in order to have a delta function interaction and only one nonlinear term. When we consider higher densities, the ground state energy changes, and such changes are interpreted as quantum fluctuations . Following the  series expansion of the energy per density squared in terms of $(a^3\rho)^{1/2}$ from references \cite{eigenv-Lee1957,Huang1957-hardsphere,Lee1957-manybody}, where $a$ is the two-body scattering parameter and $\rho$ the condensate density, we include two fluctuation terms in a modified Gross–Pitaevskii equation (mGPE). Additionally, we consider matter waves where the effect of three body losses and the compensating atomic feeding may be present. The loss term arises from the molecule formation inside the BEC which reduces the number of atoms occupying the ground state, and it is related to the density and the associated three body correlation function \cite{Kim2004-threebodyloss}. A counteracting feeding is necessary to balance the condensate and avoid dissipation \cite{Kagan1998-threeloss-negat-scatt}.

These systems are commonly placed in optical and magnetic traps that distort the condensate configuration. By counter propagating two laser beams with the same wavelength an optical lattice is formed. By combining optical and electromagnetic traps interesting effects may arise by choosing the right potential. Effects of resonance between BECs and the potential frequencies have been predicted and observed in optical systems. In particular, double parametric resonances have been predicted in BECs placed in oscillating parabolic traps Ref. \cite{Malomed}. Here we consider a double-well oscillating potential, described by a polynomial which includes quartic, cubic and quadratic powers of the spatial variable, and we study the effects of introducing time-dependent coefficients in front of the quadratic and cubic terms.

We will begin by studying the BEC's behavior when the double-well potential is static, and afterwards we will study how this behavior is modified when the potential barrier between the two wells changes in time. In the case of a static potential barrier, we will see that a solitary BEC's wave undergoes a ``fragmentation-regeneration" process when it passes over the potential barrier. In other words: the wave is fragmented when it passes over the barrier, but the fragments reunite again when the pulse has already passed the barrier. Then we will see that the shape of the wave is severely distorted if the potential barrier changes in time. These results would seem to imply that a high-density BEC's wave cannot survive repeated crossings over a time-dependent potential barrier. However, the distortion observed occurs when the time-dependent trapping potential is described by a function of the form $\alpha x^4+\beta(t)x^2$, but time-dependent double-well potentials can be described using other functions. Consequently, we will study the BEC's behavior when we consider a potential of the form $\alpha x^4+\beta x^2+\eta (t) x^3$ (where $\beta$ is now a constant). We will show that in this case the BEC's wave survives repeated crossings over the potential barrier. We will also see that during these crossings the BEC's wave is fragmented and regenerated, as it occurs in the case of a static potential.


This work is structured as follows. In Sec. 2 we present the modified Gross-Pitaevskii (mGP) equation that is studied in this article. We include in this mGP equation all the terms that will be considered in this work, even though the effect of the higher-order quantum fluctuation will be discussed and studied until Sec. 5. In Sec. 3 we apply the modified variational method introduced by Hasegawa, in order to obtain a set of Euler-Lagrange equations which may give us an approximate description of the evolution of the BEC's pulse. In Sec. 4 we obtain direct numerical solutions of the mGP equation. These results will show the occurrence of the fragmentation-regeneration (FR) process mentioned above. We will show that the introduction of a time-dependent coefficient in front of the quadratic term of the potential is detrimental for the pulse. Then, we will show that if the time dependence is introduced in a cubic term, the pulse will be able to cross over the potential barrier without being destroyed. To end this section, we will see that if the pulse remains within one of the potential wells, the trajectories of the pulse obtained from the numerical solution of the mGP equation and the variational equations practically coincide. In Sec. 5 we will study how the quantum fluctuation terms affect the FR process. In Sec. 6 we will take into account that the shape of the pulses observed in the numerical solutions (before and after the fragmentations) is substantially different from a Gaussian profile. Therefore, we will obtain new variational solutions using a supergaussian trial function (whose shape is closer to the pulses observed in the numerical solutions), in order to see if the change of the trial function produces a significant change in the variational results. Finally, in Sec. 7 we present a summary and the conclusions of this work.

\newpage

\section{Model}

Let us consider a dilute BEC with corrections for higher densities with atomic feeding, three- body recombination and an external trapping potential. We thus start with a mGPE of the form,
\begin{align}
\begin{aligned} i \hbar \frac{\partial \Psi(\mathbf{r}, t)}{\partial t}=&-\frac{\hbar^{2}}{2 m} \nabla^{2} \Psi(\mathbf{r}, t)+g|\Psi(\mathbf{r}, t)|^{2} \Psi(\mathbf{r}, t) \\ &+g_1|\Psi(\mathbf{r}, t)|^{3} \Psi(\mathbf{r}, t) +g_2|\Psi(\mathbf{r}, t)|^{4} \Psi(\mathbf{r}, t) \\ &-\frac{i \hbar}{2} K_{3}|\Psi(\mathbf{r}, t)|^{4} \Psi(\mathbf{r}, t)+\frac{i \gamma}{2} \Psi(\mathbf{r}, t) \\ &+V(\textbf{r},t) \Psi(\mathbf{r}, t)
\end{aligned} \label{GP-3D}
\end{align}
where $\Psi(\mathbf{r}, t)$ is the order parameter or condensate wave function, $m $ is the atomic mass, $g=\frac{4\pi \hbar^2 a}{m}$ is the two body atomic interaction constant which we consider attractive i.e. $a<0$, $g_1$ takes into account the first order Lee, Huang, Yang \cite{Lee1957-manybody} correction due to quantum fluctuations, and we take into account the next order correction from quantum fluctuations in the term characterized by the $g_2$ coefficient. The reason for introducing the next order fluctuation will be discussed in Sec. 5 in detail. $K_3$ is the real three body recombination loss coefficient, $\gamma$ the real atomic feeding parameter and $V(\textbf{r},t)$ the external trapping potential. Typical values for the mass, loss and feeding in $^{85}\text{Rb}$, are $m=1.41\times 10^{-25}\text{kg}$, $K_3=2\times10^{-26}\text{cm}^6\text{s}^{-1}$, $\gamma=3.4\times10^{-34}\text{cm}^6\text{s}^{-1}$ \cite{Altin-K3,kim2004,k3-gamm-val}. We note here that since the feeding is a controlled process it can be tuned to higher values as desired. Moreover, the three-body recombination loss may have a greater magnitude \cite{Abdullaev-K3}, and be taken to depend quadratically on the scattering length, \cite{Altin-K3,Boesten-K3} $K_3=2.68\times10^{-13}a^2\text{cm}^4\text{s}^{-1}$. If the scattering length is tuned, as in our case, this dependence promotes consistent numerical solutions.

The origin of the quantum fluctuation terms in \eqref{GP-3D} can be traced back to the non-delta function interaction between particles in the BEC, which in turn modifies the ground state energy. This change must depend on the density and the strength of the interaction, since for denser gases more collisions take place, thus the form of the interaction potential becomes more important. This turns out to be correct, see Refs.\cite{eigenv-Lee1957,Huang1957-hardsphere,Lee1957-manybody} where the pseudopotentials technique is used to calculate changes in the energy per particle as a series expansion in terms of $a$, the two-body scattering parameter, and $n=|\Psi|^2$ the condensate density. To introduce these new contributions, we consider the terms as ``quantum fluctuations" from the ground state energy which in the low-density regime has the form \cite{kim2004}:
\begin{align}
\begin{aligned}
\epsilon(n)=& \frac{2 \pi \hbar^{2}}{m} a n\left[1+\frac{128}{15 \sqrt{\pi}}\left(n a^{3}\right)^{1 / 2}\right.\\
&\left.+8\left(\frac{4 \pi}{3}-\sqrt{3}\right) n a^{3}\left[\ln \left(n a^{3}\right)+C\right]+\cdots\right] 
\end{aligned} \label{GS-Energy}
\end{align}
where $\epsilon(n)$ is the  ground state energy per particle and the corrections are introduced into a Kohn-Sham (KS)-like equation \cite{kim2004} by adding $\frac{\partial(n \epsilon(n))}{\partial n}$. In this way all nonlinear terms, including the $g$ term, of the mGPE considered before are derived. For higher densities this equation becomes inaccurate and a new ground energy expression is needed, as shall be discussed in Sec.5. In equation \eqref{GP-3D} we took the first, second and third terms, not including the logarithm since it was seen to give wrong results in ref. \cite{log-term}, while the second and third term represent a significant improvement over the mean field approximation \cite{kim2004,log-term,braaten2002}. From the expansion \ref{GS-Energy} we can also estimate the coefficients $g_1,g_2$ which turn out to be \cite{braaten2002}: $g_1=\frac{128\sqrt{\pi} \hbar^2 a^{5/2}}{3 m}$, $g_2\approx\frac{600\pi \hbar^2 a^{4}}{3 m}$.

Next, we consider an effective 1-D form of equation \eqref{GP-3D}, with $\Psi(\textbf{r},\tau)=\Phi(r)u(z,\tau)$ in cylindrical coordinates, by considering axial symmetry and averaging over the radial density profile, which we suppose takes the stationary form of the GPE, a gaussian with standard deviation \cite{Abdullaev-a-perp} $a_\perp\approx2\times10^{-6}\text{m}$ and amplitude $1/\sqrt{\pi} a_\perp$. After the integration we seek to simplify the numerical and analytical analysis by expressing the 1-D equation in a dimensionless form so we change variables to $z=(a_\perp/\sqrt{2})x, \; t=\omega_\perp \tau, A=\sqrt{2\pi\hbar^2/|g|m}$, where $A$ is a normalization factor such that $u=A\psi(x,t)$, and $\omega_\perp$ is the radial frequency defined by $\omega_\perp=\hbar/ma^2_\perp$ which essentially factors the dimensions out of $V(\textbf{r},\tau)$. These changes not only make the 1-D equation dimensionless in the coefficients and variables, they also allow us to have the time and space derivatives, the $g$ non-linear term and the potential coefficients normalized to one. With these changes the 1-D dimensionless form of the mGPE turns out to be:
\begin{align}
\begin{aligned} i \frac{\partial \psi}{\partial t}=&-\pdv[2]{ \psi}{x}+V(x,t)\psi-|\psi|^{2} \psi +g_1|\psi|^{3} \psi  \\&+g_2|\psi|^{4}\psi-\frac{i K_{3}}{2} |\psi|^{4} \psi+\frac{i \gamma}{2} \psi
\end{aligned} 
\label{1-D-adim-mGPE}
\end{align}
where we have relabeled the dimensionless 1-D coefficients as $g_1,g_2,K_3,\gamma$ for convenience. Their values are now given by $|g_1|=\frac{64\sqrt{2}|a|}{15\pi a_\perp}, |g_2|=\frac{50a^2}{\pi a^2_\perp},K_3=\frac{m K'_3}{12\pi^2a^2a^2_\perp\hbar},\gamma=\frac{\gamma'}{\hbar \omega_\perp}$, which are manifestly dimensionless and the primes denote now the 3-D original values. The quantum fluctuation terms can have a wide range of values depending on the scattering and transverse lengths. Experimentally the best way to modify the coefficient values is by Feshbach resonances \cite{Feshbach-a-tune-2000,Feshbach-a-tune-1999,Feshbach-PRA}  that can tune the value of the scattering length $a$ at will, a common example is a $^{85}\text{Rb}$ condensate  where the change can even turn $a$ from repulsive to attractive, and vice versa. Taking this scattering length to be 10 times that of  typical $^{85}\text{Rb}$ measurements \cite{Rb-85-scatt-a} we find that the dimensionless coefficients take the values $|g_1|=0.1$, and $|g_2|=0.05$. With this value of $a$ we obtain the dimensionless $K_3\approx0.001$, and we take $gamma$ similar to counterbalance the loss.
Concerning the form of the function $V(x,t)$, it is worth remembering that harmonic potentials are frequently used to confine BECs. In such cases the behavior of BECs is rather simple: either they stay motionless at the bottom of the potential, or they oscillate periodically.
However, the behavior of BECs and Fermi superfluid’s confined in double-well potentials is more interesting \cite{Adhikari-1,Adhikari-2}. Consequently, in the present paper we investigate the behavior of a dilute BEC  with higher density and quantum corrections placed in a time-varying double-well potential containing quartic, cubic and quadratic terms. The quadratic term defines the potential barrier that separates the two wells of the potential. To consider a harmonic time dependence we could modify the quadratic coefficient or the cubic one. As we shall see in Sec. 4 a time-dependent quadratic term destroys the pulse, thus only the cubic term promotes the BEC's regeneration and shall be the focus of the next section. In summary, the potential that we shall consider here in general is:
\begin{align}
V(x,t)=\alpha x^4-\beta(t) x^2+\eta \sin(\omega_0 t)x^3, \label{potent}
\end{align}
modelling a double-well oscillating potential. 
The wave function  is normalized to the number of particles in the condensate so that the dimensionless version is taken to be ${\displaystyle\int_{-\infty}^{+\infty}|\psi|^{2} \mathrm{d} x=N }$. 

\section{Variational Approximation}
In this section we will obtain approximate solutions of Eq. \eqref{1-D-adim-mGPE} using solitary pulses as initial conditions. We will propose a tentative solution which depends on different pulse parameters such as height, width, position, phase and chirp, all of which depend on time. There exist different ways of determining the temporal evolution of these parameters. Some methods are based on the conserved quantities of the equation under study \cite{Biswas-1, Biswas-2}, others use the eigenfunctions of the linearized equation \cite{Hoseini}, and there are variational methods which can be applied when the equation under study (or a part of it) can be obtained from a suitable Lagrangian density \cite{ Hasegawa,VA-Anderson, Carretero}. In the present article we will use a variational method, as these methods have been very successful to obtain approximate solutions of nonlinear equations similar to Eq. \eqref{1-D-adim-mGPE}, which describe the propagation of light pulses in optical fibers and liquid crystals \cite{JF1,JF2,JF3,JF4,JF5}.

Equation \eqref{1-D-adim-mGPE} cannot be obtained from a Lagrangian density, and therefore the traditional method of Anderson \cite{VA-Anderson} cannot be applied to this equation. However, if we rewrite this equation in the form:
\begin{align}
\begin{aligned} i \psi_t&+\psi_{xx}+|\psi|^{2} \psi -g_1|\psi|^{3} \psi -g_2|\psi|^{4} \psi \\ &-\left[\alpha x^4-\beta x^2+\eta \sin(\omega_0 t)x^3\right] \psi= i R(\psi,\psi^*)
\end{aligned} 
\label{THEequation_2}
\end{align}
where $R(\psi,\psi^*)=-\frac{1}{2} K_{3}|\psi|^{4} \psi+\frac{\gamma}{2} \psi$, we can see that if we drop the term $R(\psi,\psi^*)$ the resulting equation can be obtained from the Lagrangian density:
\begin{align}
\begin{aligned} \mathcal{L}=& 
\frac{i}{2}\left(\psi^{*} \psi_t-\psi \psi_t^*\right)+\dfrac{1}{2} |\psi|^4-\frac{2g_1}{5} |\psi|^5-\frac{g_2}{3}|\psi|^6\\ &-\left[\alpha x^4-\beta x^2+\eta \sin(\omega_0 t)x^3\right]|\psi|^2-|\psi_x|^2
\end{aligned}
\label{lagrangian}
\end{align}
and therefore Eq. \eqref{THEequation_2} can be written as a modified Euler-Lagrange equation of the form:
\begin{align}
\fdv{\mathcal{L}}{\psi^*}=\pdv{\mathcal{L}}{\psi^*}-\pdv{t}\pdv{\mathcal{L}}{\psi_t^*}-\pdv{x}\pdv{\mathcal{L}}{\psi_x^*}=i R(\psi,\psi^*)
\end{align}
For equations of this form, we can use the modified variational method presented by Hasegawa in Ref. \cite{Hasegawa}, in order to calculate the time evolution of the pulse parameters which appear in the proposed trial function. 
Taking into account that the equilibrium distribution of an ideal non-interacting condensate has a gaussian nature, we will use a trial function of the form:
\begin{align}
\psi(x,t)=A(t)\exp \left\{-\dfrac{(x-c(t))^2}{2a^2(t)}+i\left( b(t)x^2+f(t)x+h(t)\right) \right\}\label{Gauss-2}
\end{align}
where the functions $A,a,b,c,h$ and $f$ are the time dependent parameters mentioned above. A different trial function will be used in Sec. 6. If these six parameters are denoted as  $r_n(t),(n=1,\ldots,6)$ , the evolution of these functions will be defined by a system of equations of the form \cite{Carretero}:
\begin{align}
\frac{\partial L}{\partial r_{n}}-\frac{d}{d t} \frac{\partial L}{\partial \dot{r}_{n}}=-i \int\left(R^{*} \frac{\partial \psi}{\partial r_{n}}-R \frac{\partial \psi^{*}}{\partial r_{n}}\right) d x.\label{VA}
\end{align}
where ${\displaystyle L(r_n(t))= \int_{-\infty}^{\infty}\mathcal{L}(r_n(t);x) \; dx}$  is the so-called averaged Lagrangian. In the present case, the integration of the Lagrangian density (5) leads to an averaged Lagrangian of the following form:

\begin{align}
&L=-\frac{3}{4} \sqrt{\pi } a^5 \alpha  A^2-2 \sqrt{\pi } a^3 A^2 b^2+\frac{1}{2} \sqrt{\pi
} a^3 A^2 \beta -\frac{1}{2} \sqrt{\pi } a^3 A^2 \dot{b}\notag\\
&-3 \sqrt{\pi } a^3 \alpha  A^2
c^2-\frac{3}{2} \sqrt{\pi } a^3 A^2 c \eta  \sin \left(t \omega _0\right)-4 \sqrt{\pi
} a A^2 b^2 c^2\notag\\
&-4 \sqrt{\pi } a A^2 b c f-\sqrt{\pi } a
\alpha  A^2 c^4+\sqrt{\pi } a A^2 \beta  c^2\\
&-\sqrt{\pi } a A^2 c^3 \eta  \sin \left(t
\omega _0\right)-\sqrt{\pi } a A^2 c \dot{f}-\sqrt{\pi } a A^2 f^2-\frac{1}{3}
\sqrt{\frac{ \pi }{3}} a A^6 g_2\notag\\
&-\frac{2}{5} \sqrt{\frac{2 \pi }{5}} a A^5
g_1+\frac{1}{2} \sqrt{\frac{\pi }{2}} a A^4 -\sqrt{\pi } a A^2
\dot{h}-\frac{\sqrt{\pi } A^2}{2 a}-\sqrt{\pi } a A^2 \dot{b} c^2 \notag
\end{align}
Substituting this Lagrangian in Eq. \eqref{VA} we obtain a system of six coupled differential equations which, after some algebraic rearrangements, take the following form:
\begin{align}
& \dot{A}=-\frac{1}{18} A \left(4 \sqrt{3} A^4 K_3+36 b-9 \gamma \right) \label{dot-A}\\
&\dot{a}=\frac{1}{9} a \left(\sqrt{3} A^4 K_3+36 b\right)\label{dot-a}\\
& \dot{c}=2 (2 b c+f)\label{dot-c}\\
& \dot{b}=
-3 \alpha  a^2+\frac{1}{a^2}\left( \frac{2\sqrt{3}}{9} A^4 g_2+\frac{3}{5}\sqrt{\frac{2}{5}} A^3 g_1-\frac{A^2 }{2 \sqrt{2}}\right) \label{dot-b} \\
&+\frac{1}{a^4}-4 b^2+\beta -3 c \left(2 \alpha c+\eta  \sin ( \omega_0 t)\right) \notag\\
& \dot{h}= \frac{5 A^2 }{4 \sqrt{2}}-c^3 \left(3 \alpha  c+\eta  \sin \left(\omega
_0 t\right)\right)-f^2+\frac{3 a^4 \alpha }{4}\notag\\
&+\frac{c^2}{a^4}+\frac{3}{2} a^2 c \left(2 \alpha  c+\eta  \sin \left(
\omega _0 t\right)\right)-\frac{4 A^4 g_2}{3\sqrt{3}}-\frac{13 A^3 g_1}{5
	\sqrt{10}}\label{dot-h}\\
&+\frac{c^2}{a^2} \left(\frac{2\sqrt{3}}{9} A^4
g_2+\frac{3}{5} \sqrt{\frac{2}{5}} A^3 g_1-\frac{A^2}{2 \sqrt{2}}\right)-\frac{1}{a^2} \notag\\
&\dot{f}=-\frac{3}{2} \eta  \left(a^2-2 c^2\right) \sin \left( \omega _0 t\right)-\frac{2c}{a^4}-4 b f+8 \alpha  c^3\notag\\
&-\frac{4\sqrt{3} A^4 c g_2}{9 a^2}-\sqrt{\frac{2}{5}}\frac{6  A^3 c g_1}{5a^2}+\frac{A^2 c }{\sqrt{2} a^2}, \label{dot-f}
\end{align}

\begin{figure}[H]%
	\centering
	\begin{subfigure}{.8\columnwidth}
		\includegraphics[width=\columnwidth]{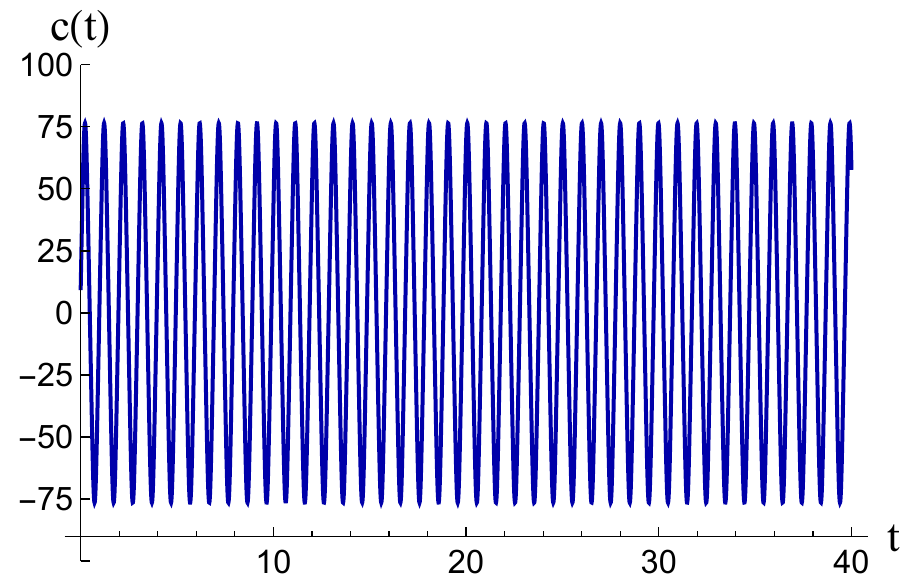}%
		\caption{}%
		\label{eta0-c}%
	\end{subfigure}\hfill%
	\begin{subfigure}{.8\columnwidth}
		\includegraphics[width=\columnwidth]{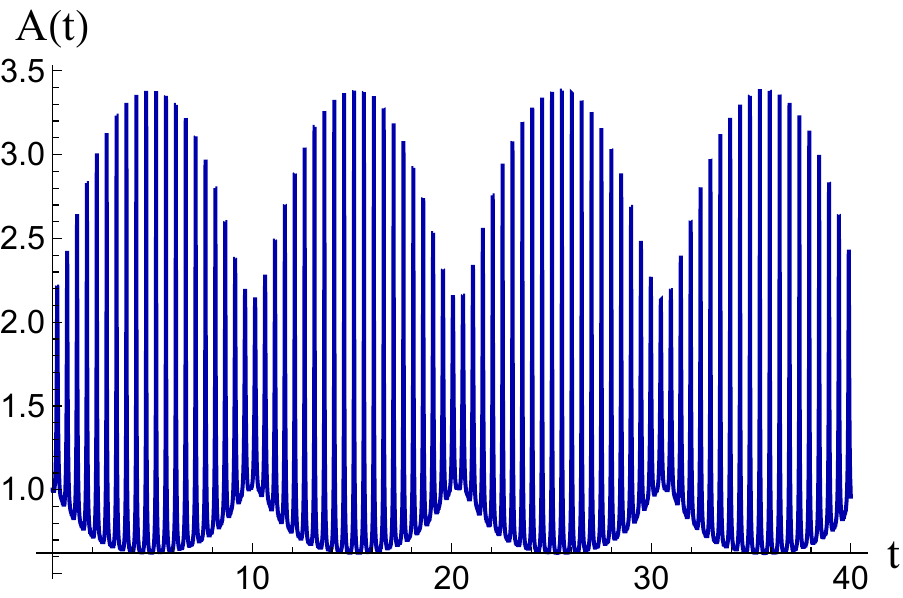}%
		\caption{}%
		\label{eta0-A}%
	\end{subfigure}\hfill%
	\begin{subfigure}{.8\columnwidth}
		\includegraphics[width=\columnwidth]{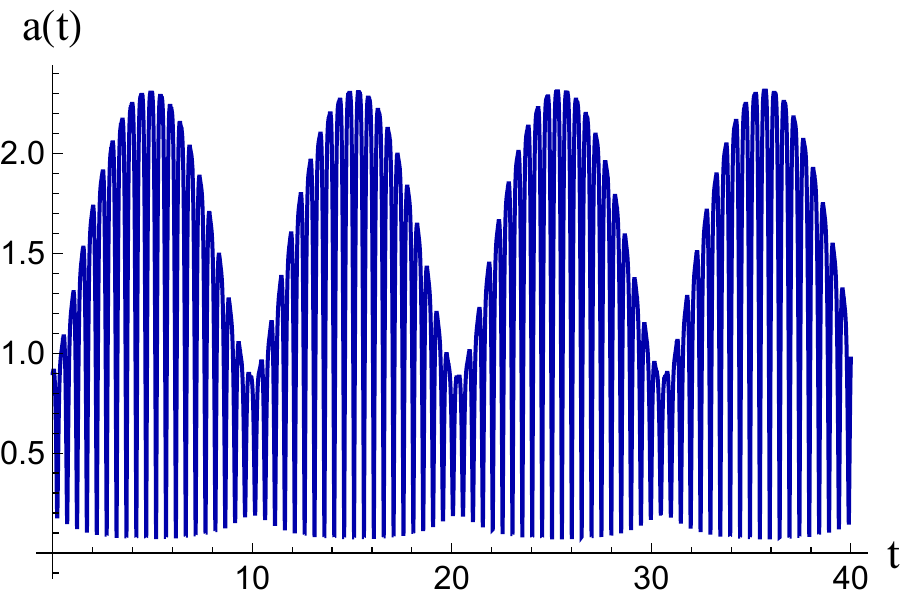}%
		\caption{}%
		\label{eta0-a}%
	\end{subfigure}\hfill%
	\begin{subfigure}{.8\columnwidth}
		\includegraphics[width=\columnwidth]{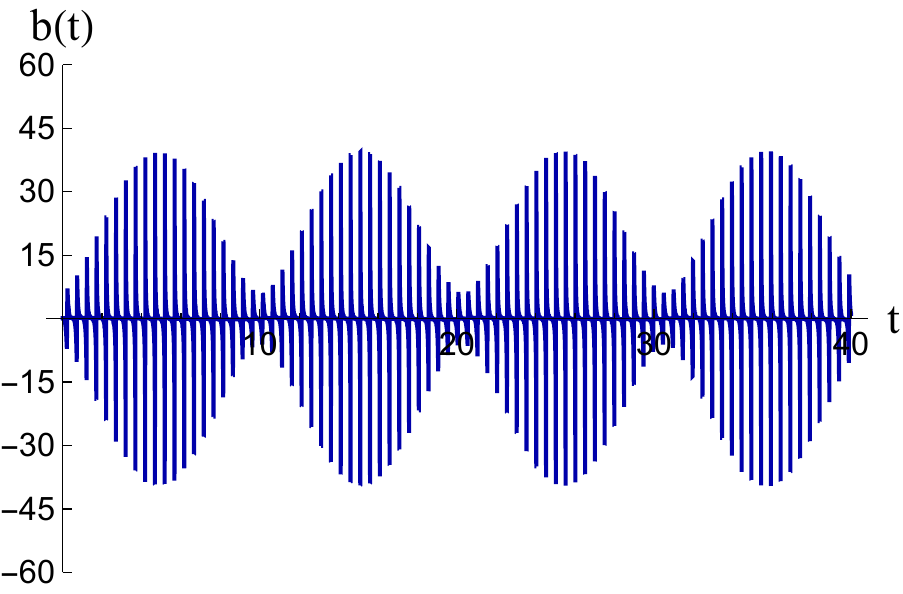}%
		\caption{}%
		\label{eta0-b}%
	\end{subfigure}%
	\centering
	\caption{Numerical solutions for the a) center of mass, b) amplitude, c) width and d) chirp of equations \eqref{dot-A}-\eqref{dot-f} with initial conditions $A(0) = 1, a(0) = 0.9,c(0) = 10, b(0) = 0.1, h(0)= -0.1, f(0) = 200$ and coefficients $\alpha = 0.0012,
		\beta = 0.05,\eta= 0,g_1 = -0.1,g_2= 0.05, K_3= 0.001, \gamma = 0.0015 ,\omega_0=\pi/10$. }
	\label{val-sol-eta0}
\end{figure}


From equations \eqref{dot-A}-\eqref{dot-f} we can obtain the stationary solution if there is no time dependence in the potential,$\eta=0$. In this case the solution for $\alpha = 0.0012,
\beta = 0.05,\eta= 0.05,g_1 = -0.1,g_2= -0.05, K_3= 0.001, \gamma = 0.0015$ is given by $A=1.27, b=-0.000125,c=7.788,a=1.004,f=0.001947,h(t)=-17.68 t$. We will use similar values of the width, amplitude and other parameters in the subsequent initial conditions so as to have more stable solutions. Now since we want the pulse to move, we will take wave number $f$ bigger than this static solution. The V.A. solutions for the complete potential \eqref{potent} will be discussed in Sec. 4 , after a discussion of the time dependent term.

If there is no time dependence, the solutions for the parameters are generally of the form of Fig. \ref{val-sol-eta0}. The center of mass oscillates steadily, as shown in Fig. \ref{eta0-c} in a sinusoidal manner. The amplitude and frequency of the motion for the center of mass seemed to vary inappreciably while the initial conditions changed, as long as the initial value of $f$ was not touched. If $f_0$ increases/decreases  the amplitude increases/decreases with it. Meanwhile the amplitude $A(t)$, Fig. \ref{eta0-A} , and width $a(t)$, Fig. \ref{eta0-a}, oscillate with a modulation whose frequency was seen to depend on the chirp initial value. For large values of the chirp all quantities oscillate faster, even the center of mass. 

This behavior shows that even though the form of the parameters has only one mode, their amplitude and oscillation  does depend strongly on the initial conditions, especially on the chirp $b$ and wave number $f$. Finally, we note that the form of the chirp , Fig. \ref{eta0-b}, as a function of time is oscillating with a modulation that has the exact same period as $a$ and $A$, thus it is controlled by the initial chirp value. The time dependence of these	 parameters seems simple as they only change in the amplitude and the frequency of the oscillations. Next, we will consider what happens when the potential changes in time \eqref{potent}.
\section{Numerical testing}

To investigate the behavior of the pulse when the potential has a time dependence, we consider three possible cases before introducing the potential of equation \eqref{potent}. We will see that only this potential, between the alternatives, can have nonzero time dependence, and non-destructive influence on the pulse. First let us consider the simple case we have just analyzed of a time independent potential \eqref{potent} with $\eta=0$, in this case the variational solutions seem to imply a stable oscillating pulse that varies periodically in amplitude, width, chirp and center of mass.

In the present section we will investigate if this simple behavior is authentic, how it is modified with different potentials and some interesting observations. In order to do so, we will obtain direct numerical solutions of Eq. \eqref{THEequation_2} using initial conditions similar to those used to obtain the variational solutions presented in the previous section. The numerical solution of Eq. \eqref{THEequation_2} corresponding to an initial condition similar to that used to obtain Fig. \ref{val-sol-eta0}, shows that the behavior of the pulse is not so simple as the variational solution seems to imply.

\begin{figure}[H]%
	\centering
	\includegraphics[width=0.8\columnwidth]{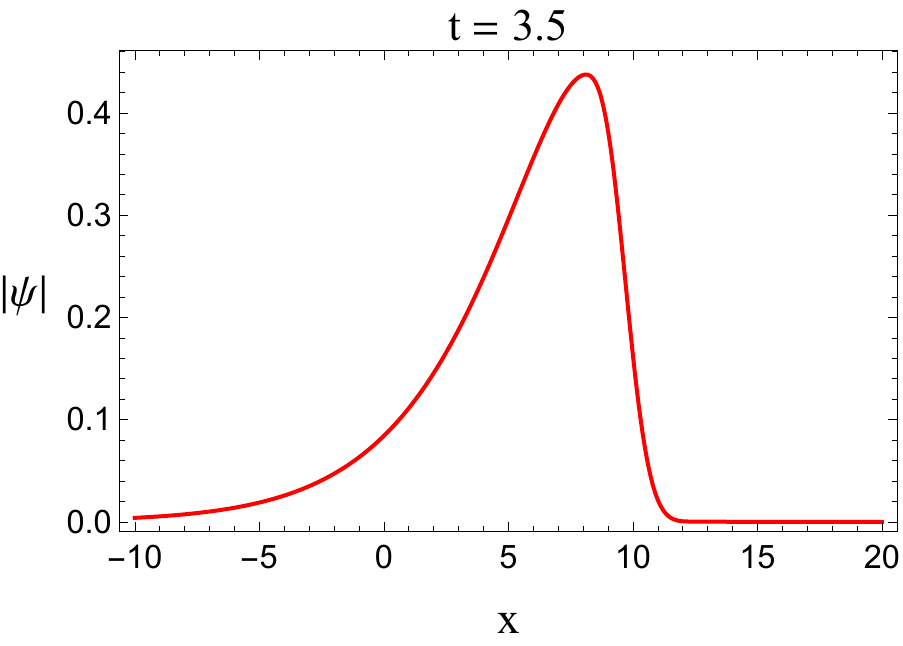}\\
	\includegraphics[width=0.8\columnwidth]{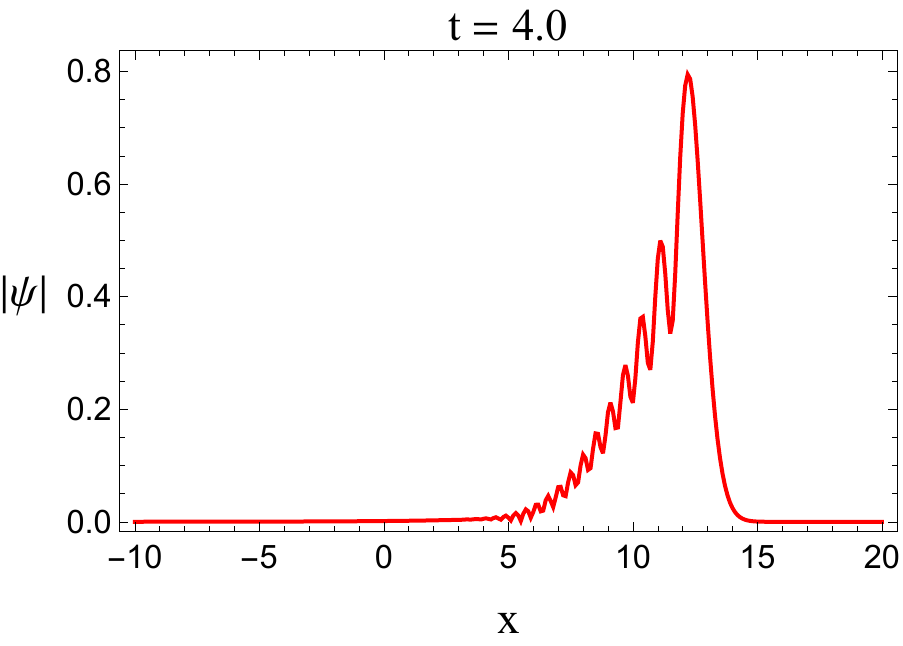}\\
	\includegraphics[width=0.8\columnwidth]{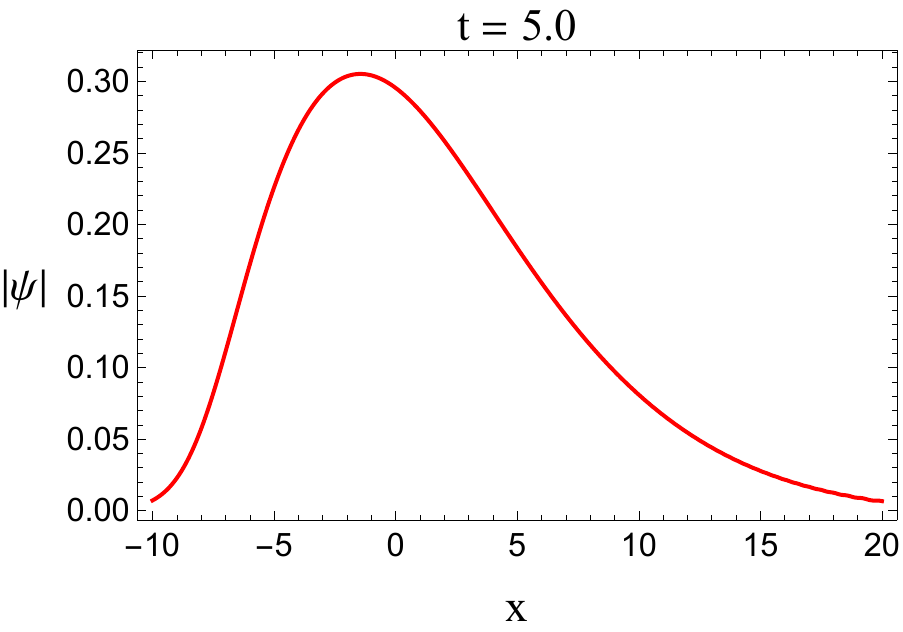}
	\caption{Numerical solutions of the Nonlinear PDE \eqref{1-D-adim-mGPE} for the soliton wave of initial condition \eqref{Gauss-2}. The values used are $A(0) = 1, a(0) = 0.5,c(0) = 10, b(0) = 0.1, h(0)= -0.1, f(0) = -100$ and coefficients $\alpha = 0.00012,\beta = 0.5,\eta= 0,g_1 = -0.1,g_2= 0.05, K_3= 0.0001, \gamma = 0.00015 ,\omega_0=2\pi/10$. }
	\label{BEC-eta0}
\end{figure}

The plots shown in Fig. \ref{val-sol-eta0} would suggest that the behavior of a BEC's pulse which obeys Eq. \eqref{THEequation_2} is rather simple, at least if the parameters $\alpha$, $\beta$ and $\eta$ have values close to those shown in this figure. Fig. \ref{eta0-c}, in particular, predicts that the pulse will just oscillate back and forth within the potential well. The numerical solution shows that the pulse does indeed oscillate (that's true), but as the pulse approaches the central potential barrier defined by the quadratic term of the potential, the pulse is fragmented into a series of narrow pulses, as shown in Fig. \ref{BEC-eta0}, at $t=4.0$. 

These narrow pulses seem to be joined to each other, a characteristic that won't survive with a time dependent potential. Once the pulse has passed completely over the potential barrier, the narrow pulses reunite again, regenerating the initial pulse. This process of fragmentation and regeneration (FR) can be observed in Fig. \ref{BEC-eta0}, where we can also observe that as the pulses reunite the width of the solution increases, and the amplitude decreases. On the other hand, let's examine what happens when we introduce time dependence in the potential.

\begin{figure}[H]%
	\centering
	\includegraphics[width=0.8\columnwidth]{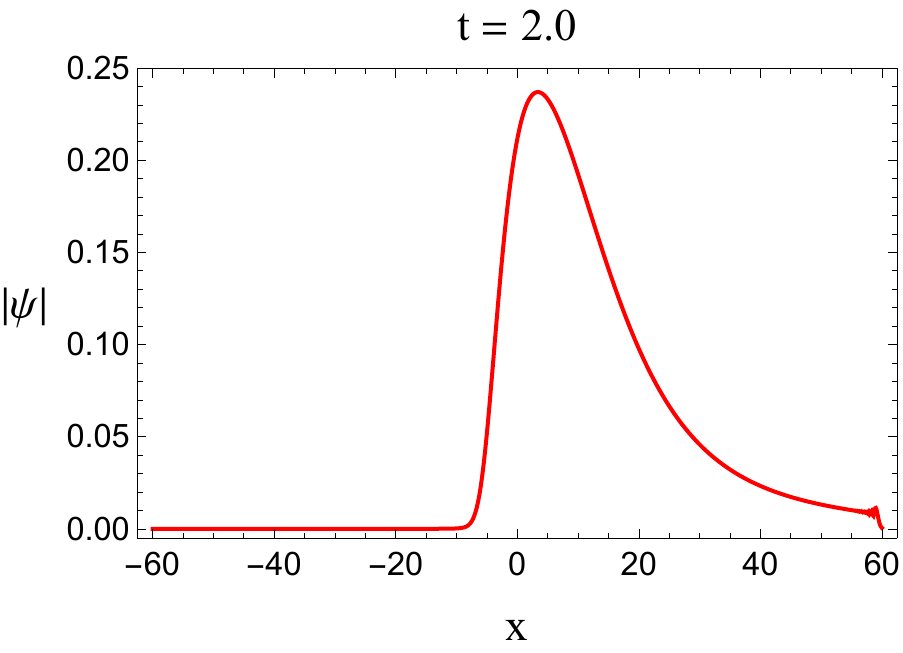}\\
	\includegraphics[width=0.8\columnwidth]{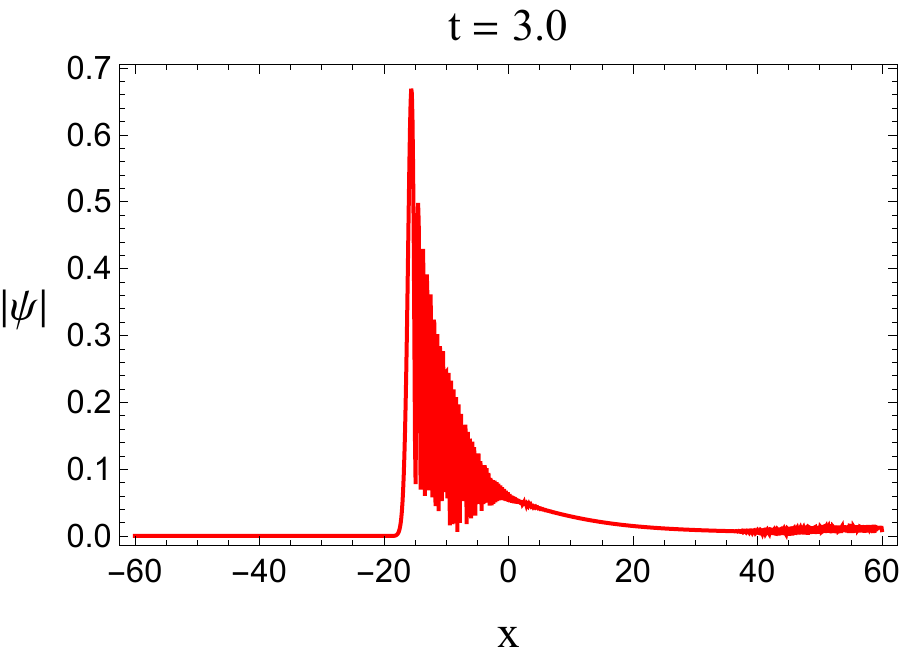}\\
	\includegraphics[width=0.8\columnwidth]{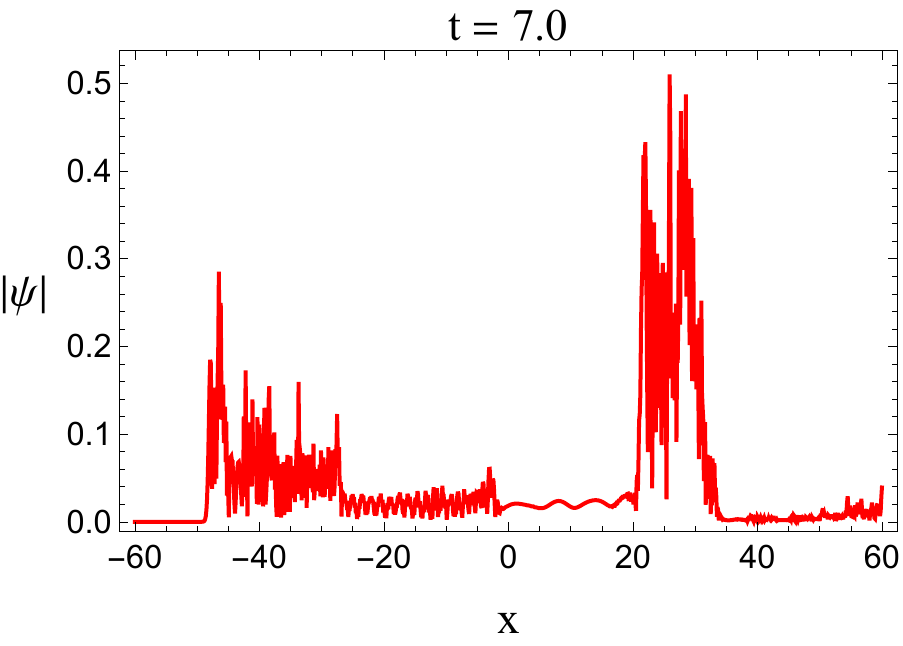}
	\caption{Numerical solutions of the Nonlinear PDE \eqref{1-D-adim-mGPE}, with harmonic time-dependence in the $x^2$ term (which destroys the pulse),  for the soliton wave of initial condition \eqref{Gauss-2}. The values used are $A(0) = 1, a(0) = 0.5,c(0) = 10, b(0) = 0.1, h(0)= -0.1, f(0) = -100$ and coefficients $\alpha = 0.00012,\beta = 0.5,\eta= 0,g_1 = -0.1,g_2= 0.05, K_3= 0.0001, \gamma = 0.00015 ,\omega_0=2\pi/10$.}
	\label{BEC-beta-sin}
\end{figure}

\begin{figure}[H]%
	\centering
	\includegraphics[width=0.8\columnwidth]{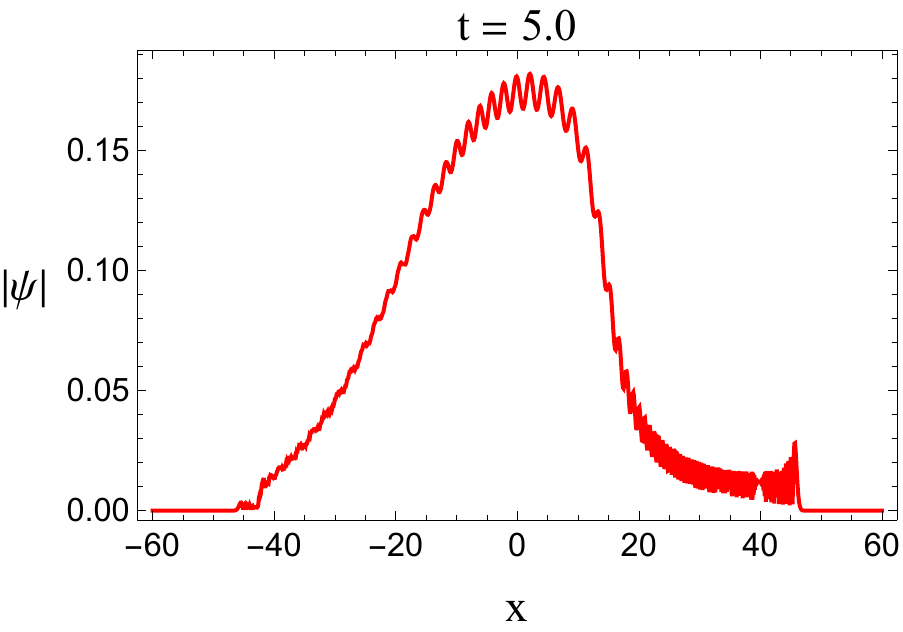}\\
	\includegraphics[width=0.8\columnwidth]{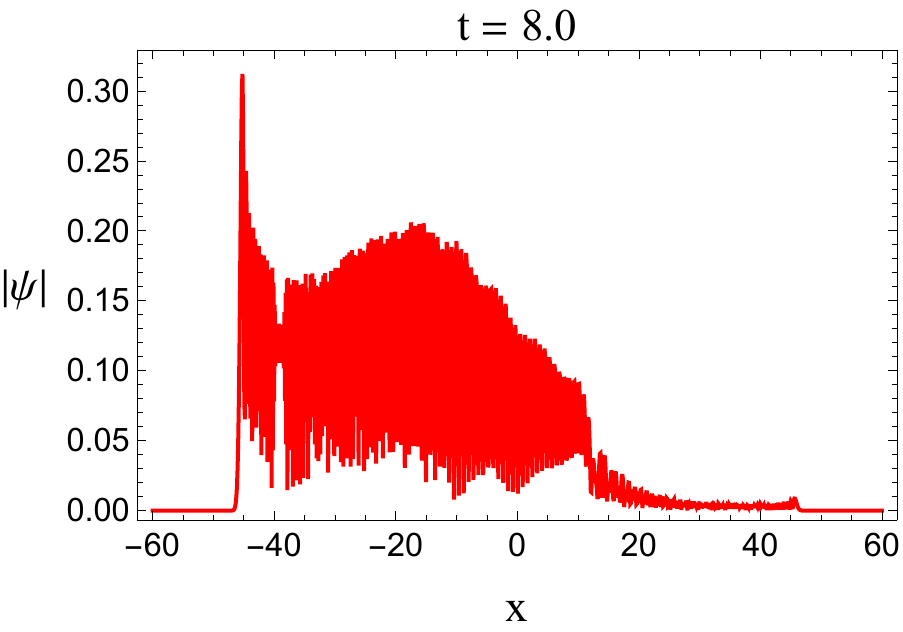}%
	\caption{Numerical solutions of the Nonlinear PDE \eqref{1-D-adim-mGPE} for the soliton wave of initial condition \eqref{Gauss-2}. A displaced harmonic time-dependence is introduced in the $x^2$ term with, $\epsilon=0.01$. The values used are $A(0) = 1, a(0) = 0.5,c(0) = 10, b(0) = 0.1, h(0)= -0.1, f(0) = -100$ and coefficients $\alpha = 0.00012,\beta = 0.5,\eta= 0,g_1 = -0.1,g_2= 0.05, K_3= 0.0001, \gamma = 0.00015 ,\omega_0=2\pi/10$. }
	\label{BEC-beta-eps}
\end{figure}

Starting from lowest order we could think of a simple movement in the $x^2$ term such that $\eta=0$ and $\beta\rightarrow\beta \sin(\omega_0 t)$. The solution profile for three different times and similar parameters as before is plotted in Fig. \ref{BEC-beta-sin}. It is clear that even though the pulse does fragment itself, as seen in Fig. \ref{BEC-beta-sin}, at $t=7.0$, the pulse's profiles at subsequent times reveal only further disintegration with no FR effect.
It appears that a time-dependent potential destroys the fragmentation and regeneration, at least if the time-dependence is introduced in the coefficient of the $x^2$ term. Even if the $\beta$ coefficient is reduced to $0.005$, numerical testing reveals that the pulse still crashes, and destroys itself. Because of this let's change the form of the time dependence, but still insisting on attaching it to the harmonic term. This time let's consider replacing the coefficient $\beta\rightarrow \beta(1+\epsilon \sin(\omega_0 t))$ where $\epsilon$ is small and positive. In this case the solution of equation \eqref{1-D-adim-mGPE} is plotted in Fig. \ref{BEC-beta-eps}.Again the pulse 

\noindent disintegrates itself as it moves and crashes against the walls of the potential. These results suggest that if a time dependent potential is to maintain the fragmentation and regeneration process, the time dependence must be introduced in a higher power of x. Therefore, we will add a cubic term in the potential, with a  time-oscillating coefficient.

\begin{figure}[H]%
	\centering
	\includegraphics[width=0.8\columnwidth]{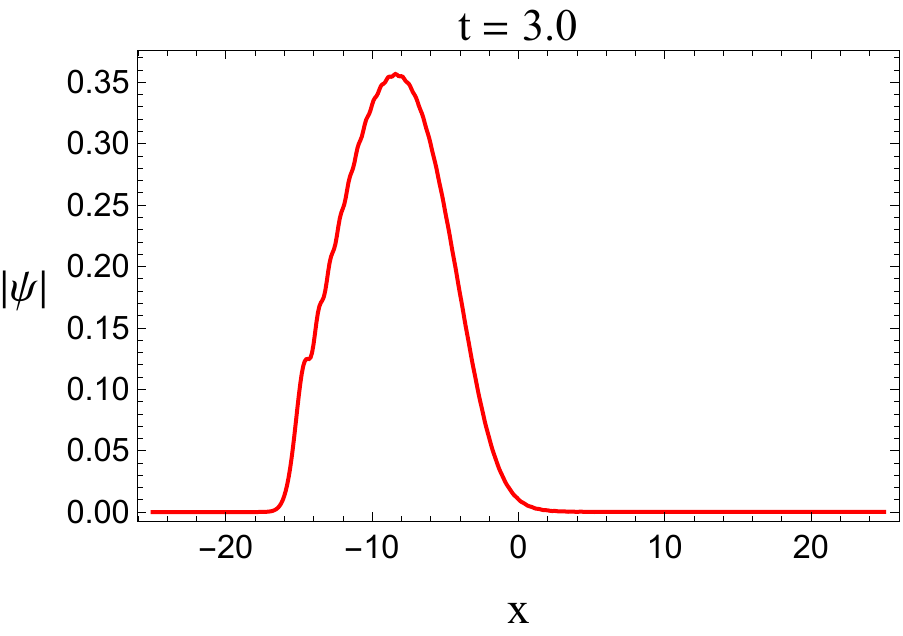}\\
	\includegraphics[width=0.8\columnwidth]{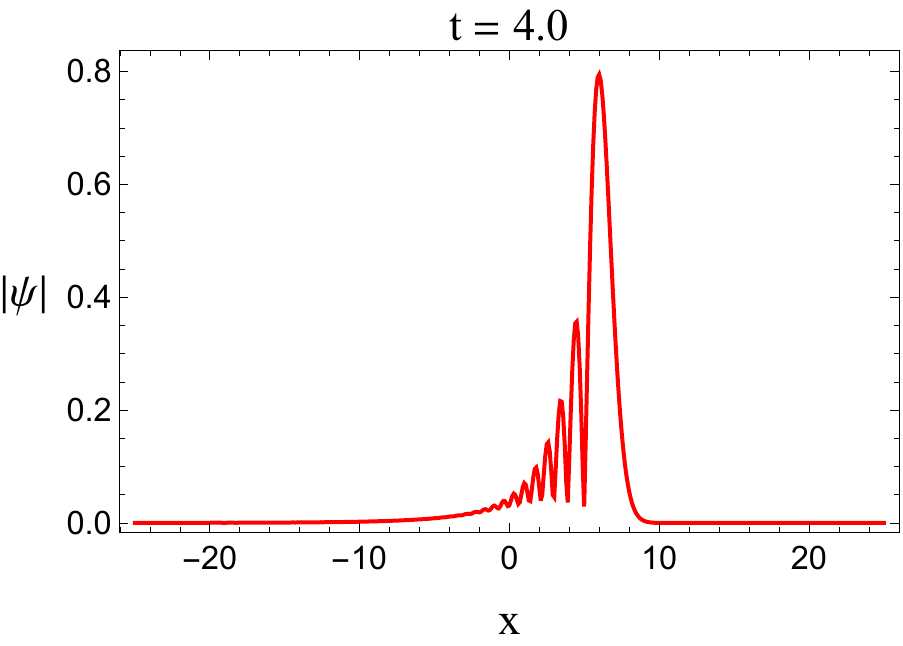}\\
	\includegraphics[width=0.8\columnwidth]{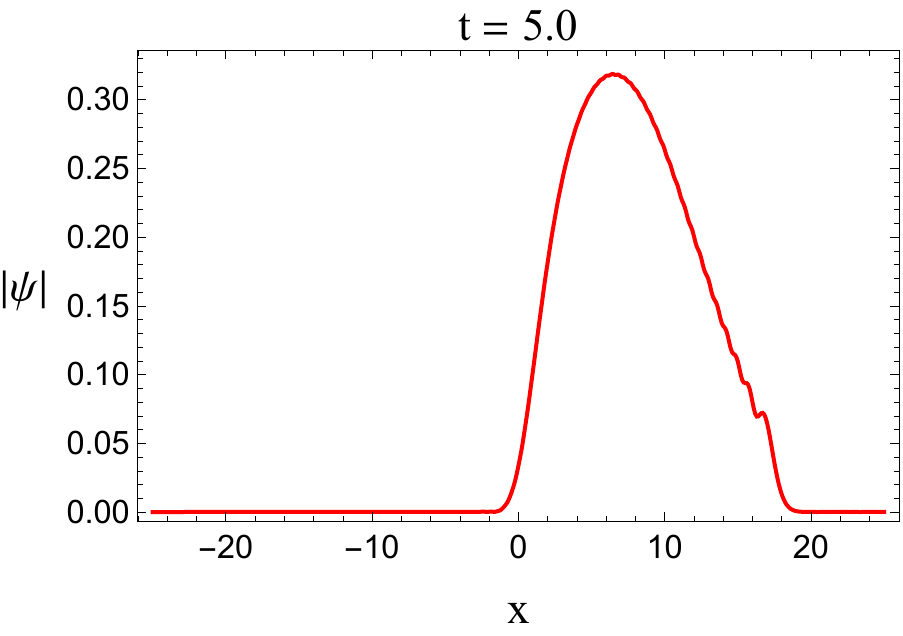}%
	\centering
	\caption{Numerical solutions of the Nonlinear PDE \eqref{1-D-adim-mGPE} for the soliton wave of initial condition \eqref{Gauss-2}. The values used are $A(0) = 1, a(0) = 0.5,c(0) = -0.01, b(0) = 0.1, h(0)= -1, f(0) = -100$ and coefficients $\alpha = 0.00012,\beta = 0.5,\eta= 0.005,g_1 = -0.1,g_2= 0.05, K_3= 0.0001, \gamma = 0.00015 ,\omega_0=2\pi/10$. }
	\label{BEC-full-potent}
\end{figure}

To corroborate this hypothesis, we analyze the authentic PDE solutions of equation \eqref{1-D-adim-mGPE} with the full-time dependent potential \eqref{potent}. The numerical solutions are shown in Fig. \ref{BEC-full-potent}. The pulse clearly fragments itself and regenerates into a single entity. Therefore, the FR process survives if the time dependence is introduced in the cubic term of the potential. Moreover, the fragmentation seen in Fig. \ref{BEC-full-potent}, at $t=4.0$, is more dramatic compared to the result for $\eta=0$ in Fig. \ref{BEC-eta0}, at $t=4.0$, with deeper valleys between pulses. In this way the potential increases the fragmentation while maintaining the regeneration of the pulse. 

\begin{figure}[H]%
	\centering
	\includegraphics[width=0.9\columnwidth]{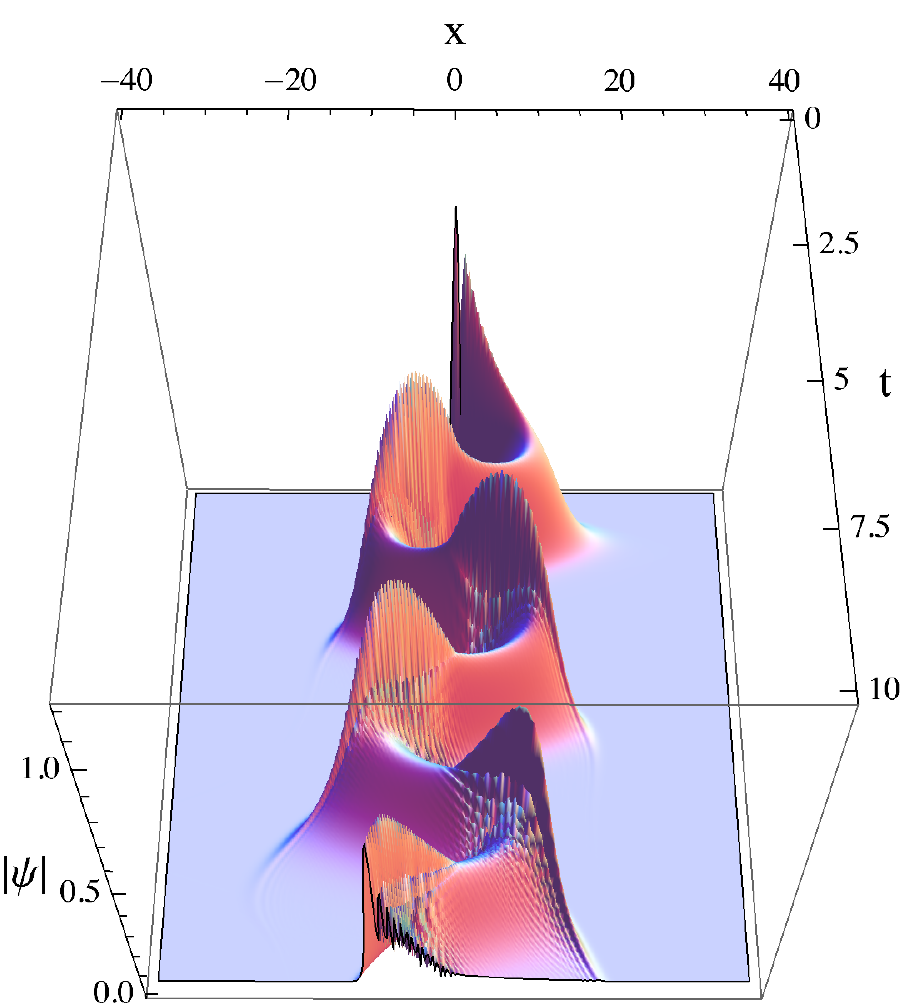}%
	\centering
	\caption{Surface plot of the numerical solutions of the Nonlinear PDE \eqref{1-D-adim-mGPE} for the soliton wave of initial condition \eqref{Gauss-2}. The coefficients and initial values are the same as in Fig. \ref{BEC-full-potent}. }
	\label{num-sol-superf}
\end{figure}

The FR process is an interesting one, as it survives still in the presence of a time dependent potential and cannot be predicted by the variational method, nor we could have imagined it in advance. However, as the fragmentation of the pulse threatens the existence of the pulse, it is reasonable to wonder if the pulse will be able to pass several times over the potential barrier 
\noindent without being destroyed. Figure \ref{num-sol-superf} gives us the answer to this question. In this figure we can see that the pulse passes once and again over the potential barrier, generating the sequence of narrow pulses during each passage, but regenerating itself in each cycle. Therefore, this periodic sequence of fragmentations and regenerations is indeed an interesting process.

\begin{figure}[H]%
	\centering
	\includegraphics[width=0.8\columnwidth]{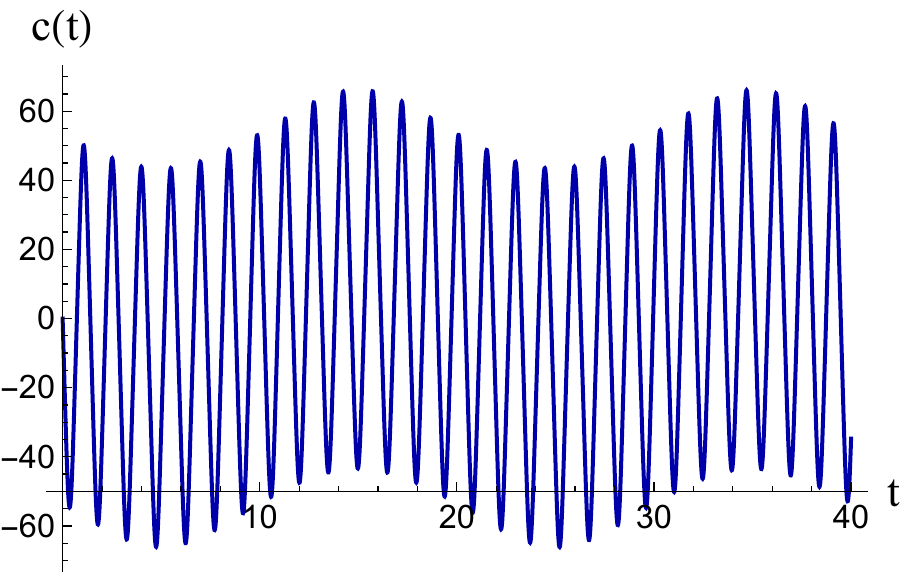}
	\caption{Numerical solution for the center of mass, equations \eqref{dot-A}-\eqref{dot-f} with initial conditions $A(0) = 1, a(0) = 0.5,c(0) = -0.01, b(0) = 0.1, h(0)= -1, f(0) = -100$ and coefficients $\alpha = 0.0012,
		\beta = 0.05,\eta= 0.05,g_1 = -0.1,g_2= 0.05, K_3= 0.001, \gamma = 0.0015 ,\omega_0=\pi/10$. }
	\label{sol-<100}
\end{figure}

\begin{figure}[H]%
	\centering
	\includegraphics[width=0.8\columnwidth]{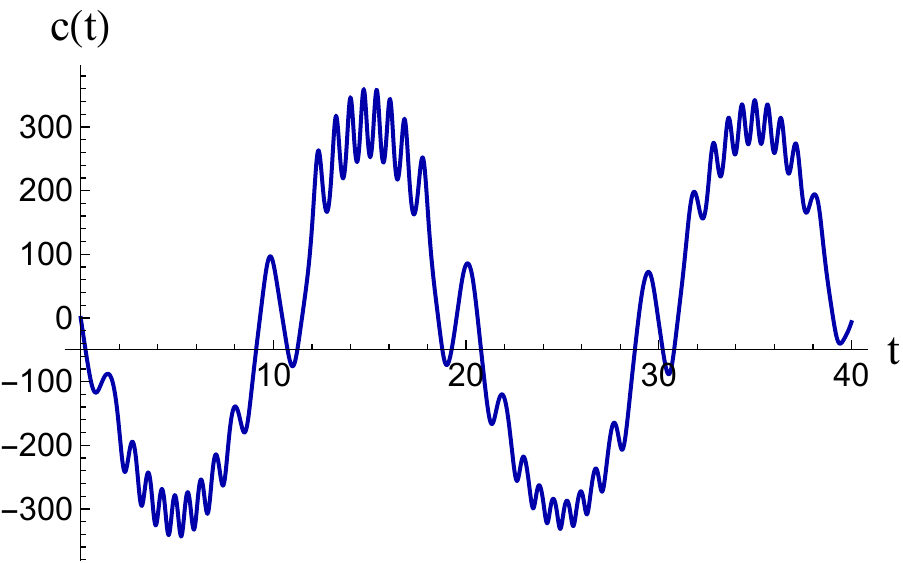}
	\caption{Numerical solutions for the center of mass of equations \eqref{dot-A}-\eqref{dot-f} with same initial conditions as Fig. \ref{sol-<100} but $\alpha = 0.00012$.}
	\label{sol->100}
\end{figure}

As a means of consistency, it is instructive to examine the behavior predicted by the V.A. and compare it to the PDE solutions, in particular to Fig. \ref{num-sol-superf}.The numerical V.A. results reveal that the solutions of Eqs. \eqref{dot-A}-\eqref{dot-f} for $\eta\not=0$ present two distinct qualitative behaviors for $c(t)$ as shown in Fig. \ref{sol-<100} and Fig. \ref{sol->100} with the initial conditions indicated in the  captions. We can observe that for small $\alpha$ the oscillations of $c(t)$ follow the potential variations, with small fluctuations near the maxima and the minima of the curve. On the other hand, for a bigger $\alpha$, the oscillation is quicker, and has approximately the potential 
\noindent frequency as the modulating frequency, just as in the $\eta=0$ case. 

If we now compare the solution shown in Figs. \ref{sol-<100}, \ref{sol->100} with Fig. \ref{num-sol-superf} we can see that the variational solution predicts qualitatively the oscillations observed in the numerical solution. However, the prediction is not quantitatively precise, since the amplitude of the oscillations is higher in the VA solution than in the numerical one. The VA also does not present bumps or additional oscillations when the pulse changes direction. The key idea here is that the pulse starts in a position, near $c=0$ the local maximum, where the oscillatory movement of the potential reflects in acute trajectory changes. The general movement change presents with fragmentation and so does not allow for accurate variational predictions that take into account one hump pulses. It will turn out that if the pulse oscillates within one of the potential wells then the solutions will be very similar. This different initial condition will be analyzed at the end of this section. In addition to the FR process, the numerical solutions of Eq. \eqref{THEequation_2}   permit us to observe a number of interesting characteristics of these solutions. In the following we describe these observations.

Observation 1:\\
The FR process depends strongly on the $x^2$ term, since for $\beta=0.05$ the pulse rapidly goes from the initial position to either side of the barrier depending on the sign of $f(0)$, and crashes with the wall produced by the $x^4$ term at approximately $x=\pm 40$ (with the same initial conditions as Fig. \ref{sol->100}) as seen in Fig. \ref{beta-0,05}. The pulse again oscillates, but the fragments disperse, and the pulse distorts in an unpredictable manner, as shown in Fig. \ref{beta-0,05} and Fig. \ref{surf-beta-0,05}. We can understand this intuitively by noticing that the $x^2$ term acts as a velocity reducing middle barrier so that the energy of the pulse gets used in passing the barrier, and this agrees with the reduction in amplitude, mentioned earlier.

Observation 2:\\
In an analogous manner, if we increase $\eta$ or $c(0)$  with a fixed frequency (corresponding, for example, to period T=10), the pulse still oscillates, but it fragments in the process, without regeneration.
When the cubic term of the potential is increased (a bigger $\eta$), the pulse is thrown against the walls of the fourth-order potential well, crashing, and destroying itself in a similar manner as in Fig. \ref{beta-0,05} and Fig. \ref{surf-beta-0,05} .This observation is expected based on the  different magnitudes of the terms involved in Eq. \eqref{THEequation_2}. The resulting form for $|\psi|$ is essentially the same as in Fig. \ref{beta-0,05} and \ref{surf-beta-0,05}.

Observation 3:\\
The fragmentation and distortion also appear for $f(0)=0$ with the same initial conditions and form as in Fig. \ref{beta-0,05}. The pulse in an originally centered position moves from one valley into the other as expected, but in the process, it deforms, crashes against the walls of the potential wells, and breaks into fragments finally dispersing between both valleys. The dispersion of the pulse in two parts can be viewed as a direct phenomenon of the oscillating potential that quickly disperses any solitary wave if it is initially placed with zero initial velocity near the center of the $x^4$ potential well. Thus, a non-zero initial velocity plays an important role on the FR process by moving the pulse into one direction, in the process delaying the dispersion on both potential valleys, and evading the crash caused by the $x^4$ term.
\begin{figure}[H]%
	\centering
	\includegraphics[width=0.8\columnwidth]{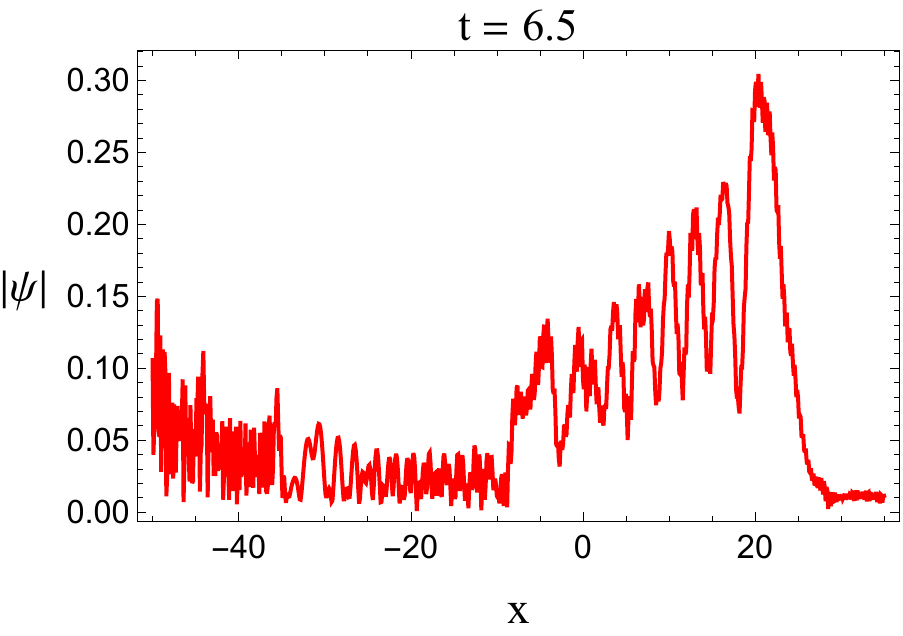}\\
	\vspace{0.5cm}
	\includegraphics[width=0.8\columnwidth]{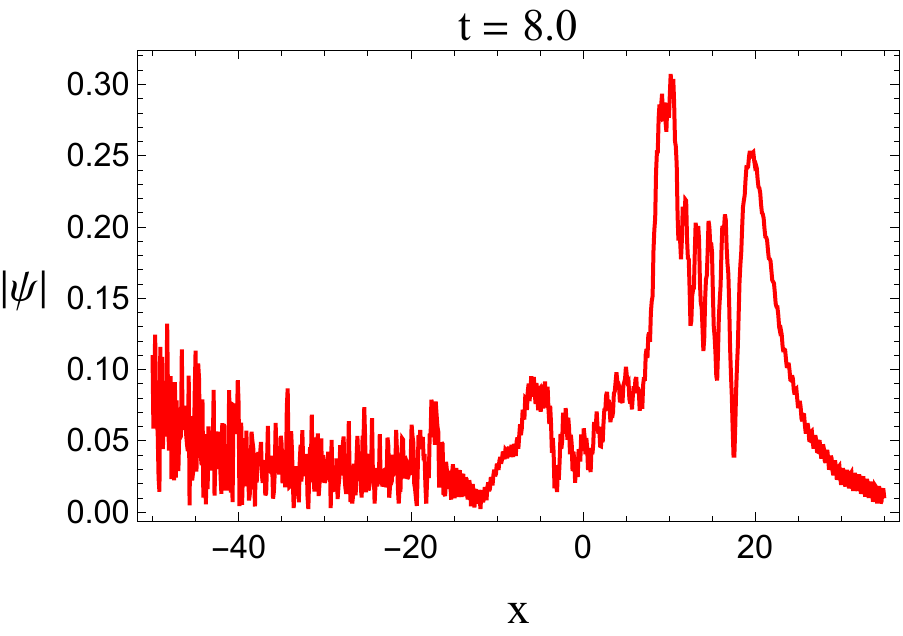}%
	\centering
	\caption{Numerical solutions of the Nonlinear PDE \eqref{1-D-adim-mGPE} for the soliton wave of initial condition \eqref{Gauss-2}.The destruction is caused by the decrease in the $\beta$ coefficient. The values used are $A(0) = 1, a(0) = 0.5,c(0) = -0.01, b(0) = 0.1, h(0)= -1, f(0) = -100$ and coefficients $\alpha = 0.00012,\beta = 0.05,\eta= 0.005,g_1 = -0.1,g_2= 0.05, K_3= 0.0001, \gamma = 0.00015 ,\omega_0=2\pi/10$.  }
	\label{beta-0,05}
\end{figure}

Observation 4:\\
On the other hand, if we fix the coefficients and initial conditions as of Fig. \ref{BEC-full-potent}  and decrease or increase the frequency (for example using $\omega_0=2\pi/100$ or $\omega_0=200\pi$ ) the behavior of the pulse does not change in a significant way, only minor shifts in the position of the pulse are observed.
In this way a frequency change does not affect greatly the behavior of the pulse, even if the frequency of the potential matches that of the pulse ($\omega_0=-2\pi/3.5$ ) . In this case the pulse maintains its height and width longer times. On the contrary, if we have minus this frequency the pulse oscillates with a bigger amplitude and broadens, reducing its movement across both valleys.

\begin{figure}[ht]%
	\centering
	\includegraphics[width=0.9\columnwidth]{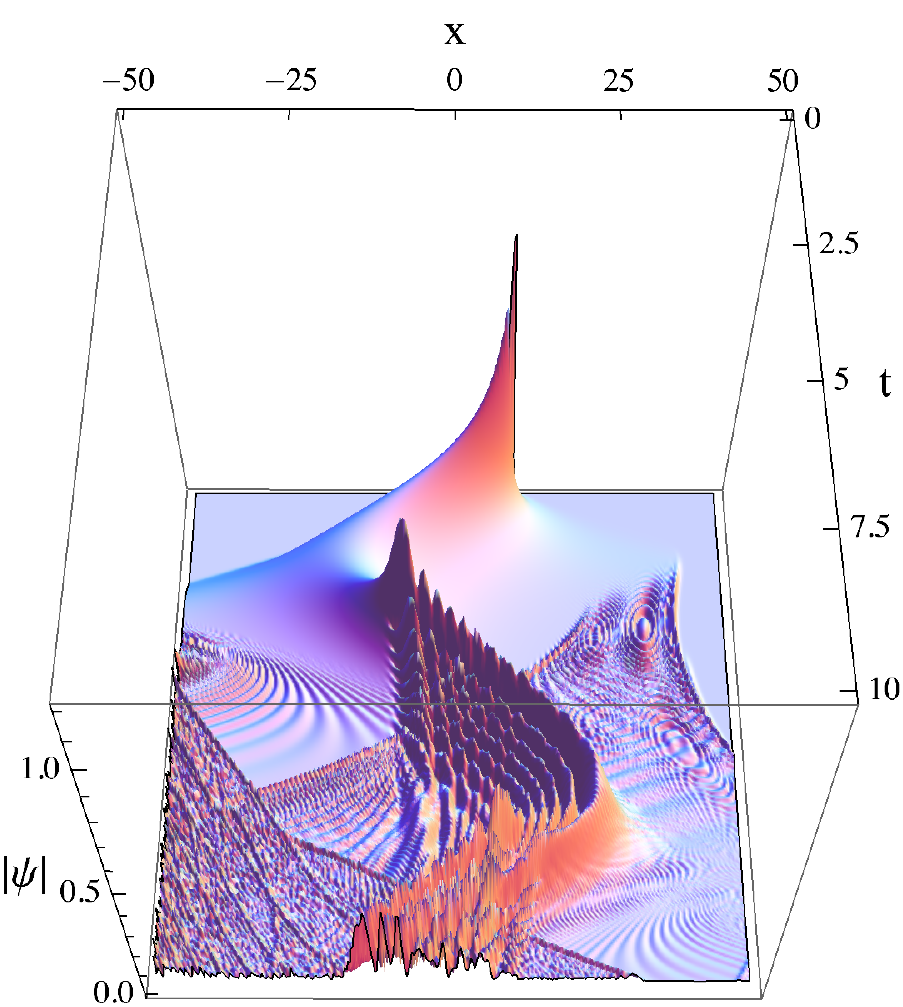}%
	\caption{Surface plot of the numerical solutions of the Nonlinear PDE \eqref{1-D-adim-mGPE} for the soliton wave of initial condition \eqref{Gauss-2}. The coefficients and initial values are the same as in Fig. \ref{beta-0,05}.}
	\label{surf-beta-0,05}
\end{figure}
\begin{figure}[H]%
	\centering
	\includegraphics[width=0.8\columnwidth]{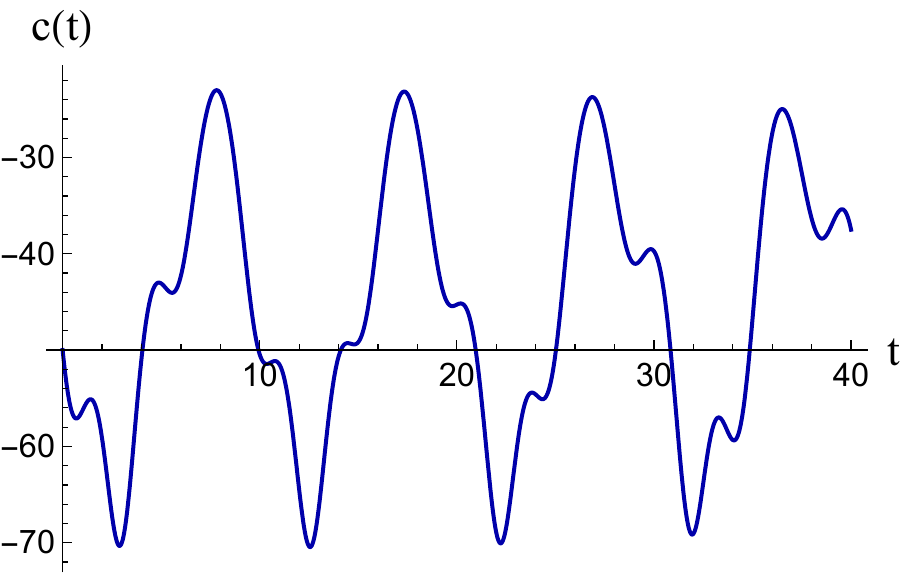}
	\caption{ Numerical solutions for the center of mass from equations \eqref{dot-A}-\eqref{dot-f} with parameters $A(0) = 1, a(0) = 0.5,c(0) = -50, b(0) = 0.1, h(0)= -1, f(0) = 0$ and coefficients $\alpha = 0.00012,\beta = 0.5,\eta= 0.005,g_1 = -0.1,g_2= 0.05, K_3= 0.0001, \gamma = 0.00015, \omega_0=2\pi/10$. }
	\label{VA-same}
\end{figure}

Observation 5:\\
If we initiate the movement of the pulse away from the center of the $x^4$ potential well ($x=0$), say $c(0)=-50$, and set zero velocity $f(0)=0$, then the pulse again presents rapid oscillations, but the center of mass $c(t)$ has a peculiar behavior. If we first use the variational method, we obtain the movement shown in Fig. \ref{VA-same}. Apart from the oscillating sine like modulation, it presents bumps at intermediate times. This behavior is directly linked to the frequency of the potential as is shown in Fig. \ref{VA-same}. The solution shown in this figure would change if we increased the period of the potential, presenting interruptions on the periodic movement, and showing additional oscillations emerging between the potential induced movement.

\begin{figure}[H]%
	\centering
	\includegraphics[width=0.9\columnwidth]{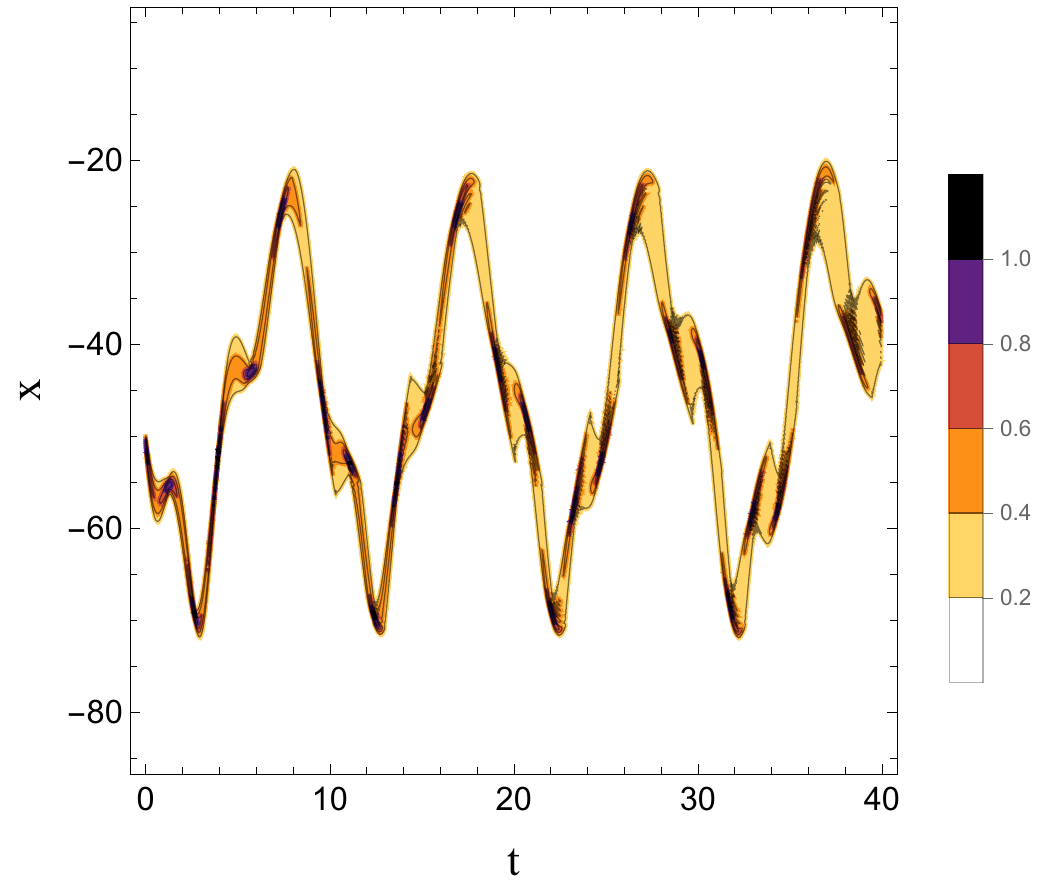}
	\caption{Contour plot of $|\psi|$ [the solution of Eq. \eqref{1-D-adim-mGPE}], for an initial condition defined by Eq. \eqref{Gauss-2} (with t=0), and the same parameters as Fig. \ref{VA-same}  }
	\label{PDE-same}
\end{figure}

If we now consider the direct numerical solution of Eq. \eqref{1-D-adim-mGPE}, again placing the BEC initially with zero velocity near the minimum of the left potential well, the pulse will start to oscillate forced by the movement of the potential. However, in this case (if we use the parameters shown in Fig. \ref{VA-same}) the pulse will not cross the potential barrier at x=0, and it will remain oscillating within the left potential well. The numerical solution of Eq. \eqref{THEequation_2} shows that the center of the pulse will move as shown in Fig. \ref{PDE-same}, if the frequency is $\omega_0=2\pi/10$. This behavior can be compared to Fig. \ref{VA-same} directly, both plots show exactly the same trajectory, even having the same changes in time and values for amplitudes of oscillation. The only difference is the slow dispersion of the pulse that reduces the maxima near turning points, thus making the intensity plot to show this range of values. Thus, by examining Figs. \ref{VA-same} and \ref{PDE-same}, we are able to realize the excellent agreement between the variational approximation (Fig. \ref{VA-same}), and the direct numerical solution of Eq. \eqref{THEequation_2}  (Fig. \ref{PDE-same}). 

\begin{figure}[H]%
	\centering
	\includegraphics[width=0.9\columnwidth]{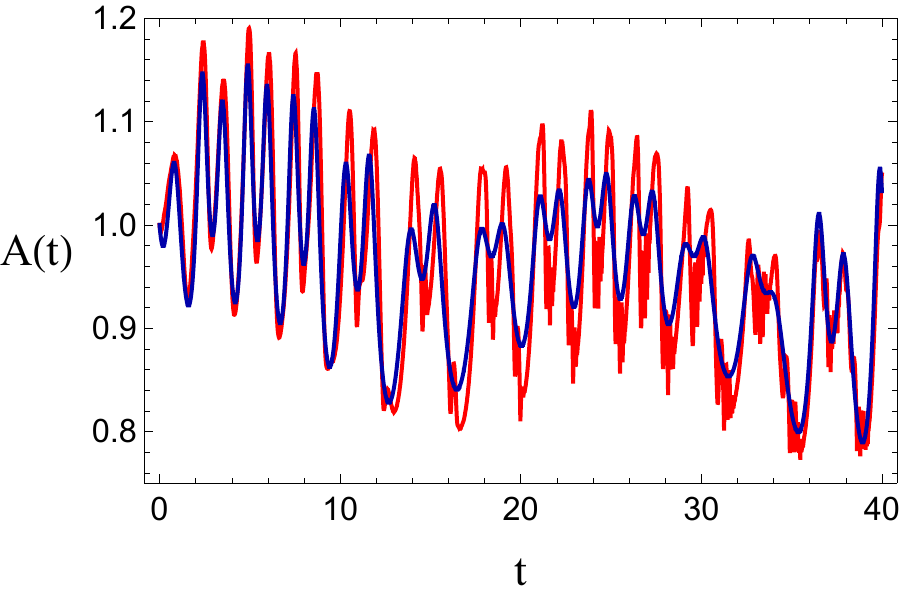}
	\caption{(Color online) Comparison of amplitudes for the VA (blue curve) and PDE (red curve) \eqref{1-D-adim-mGPE} solutions with initial conditions with parameters $A(0) = 1, a(0) = 0.9,c(0) = -50, b(0) = 0.1, h(0)= -10, f(0) = 10$ and coefficients $\alpha = 0.00012,\beta = 0.5,\eta= 0.005,g_1 = -0.6,g_2= 0.15, K_3= 0.01, \gamma = 0.0015, \omega_0=2\pi/20$ }
	\label{compar-Ampl}
\end{figure}

As a further confirmation of this agreement  consider the amplitude as a function of time predicted by the VA and the maximum of the PDE \eqref{1-D-adim-mGPE} solution, shown in Fig. \ref{compar-Ampl}. The initial conditions and  values for the coefficients are different compared to Fig. \ref{VA-same}. Because of these, one may think the agreement won't appear, but Fig. \ref{compar-Ampl} shows otherwise. The occurrence of the extrema, the shape of the oscillations, and even the values for the amplitude at the beginning and end are strikingly similar for the VA and the PDE solutions. Although  the middle region of $t\approx 13$-$35$ presents a significant difference in the precise values of both solutions, and the PDE solution has bigger oscillations, the shape and location of the maxima 


\noindent and minima are still predicted accurately by the VA. Thus, we conclude that the pulse  center and amplitude time dependence is indeed adequately  predicted by the VA. Moreover, it should be noticed that the agreement does not happen solely for the initial conditions and coefficients of Fig. \ref{VA-same}. This agreement is particularly remarkable, taken into account the complexity of the system of Euler-Lagrange equations \eqref{dot-A}-\eqref{dot-f}.
The great similarity between Figs. \ref{VA-same}, and \ref{PDE-same}, and the  comparison of solutions in Fig. \ref{compar-Ampl} proves that Anderson-Hasegawa's variational method is indeed able to describe correctly the behavior of a BEC which evolves according to Eq. \eqref{THEequation_2} , as long as the configuration of the BEC is a single-humped pulse. In this manner the oscillating behavior of the VA solutions was verified  qualitatively for the FR solutions with a strong dependence on the equation coefficients. The VA predictions are quantitatively correct when the BEC stays as a single pulse, since then the comparison of the trial function (chosen to be a single propagating pulse) matches the numerical solution of the original nonlinear PDE \eqref{THEequation_2}. Evidently, a single-humped trial function [as that shown in Eq. \eqref{Gauss-2}], is unable to reproduce the fragmentation of the pulse. This single-humpedness is the principal reason  for for the difference between the variational and the numerical solutions. In Sec. 6 we will examine another form for the trial function even more similar to the numerical solution profiles.

\section{Quantum Fluctuations}

In this section we will study the nonlinear effects coming from the quantum fluctuations of the ground state energy. We start by discussing the origin and meaning of the two terms $|\psi|^3\psi$ and $|\psi|^4\psi$. As we mentioned in Sec. 2, these terms come from the form of the interaction potential between atoms in the BEC phase. The expansion of the energy in the low-density limit \eqref{GS-Energy} results in the previous terms when introduced into the GPE. These two terms are identified as higher order terms, and are acceptable under certain assumptions. First of all, the BEC needs to be a dilute gas so that the expansion in orders of $na^3$ converges. Although the expansion allows for slightly higher densities, it does not converge in the limit $na^3 \gg 1$. The case of $|a|\rightarrow \infty$ corresponds to the unitary limit \cite{Adhikari-Unitarity, Adhikari-Superfluid}  where the use of a power series \eqref{GS-Energy} for the ground state energy leads to diverging results. The correct modification to the GPE for high densities or scattering lengths is via Padé approximants \cite{Adhikari-Unitarity, Adhikari-Superfluid} which allow to write the chemical potential in a non-perturbative expression with complete convergence from the weak coupling to unitary limits.

\begin{figure}[H]%
	\centering
	\begin{subfigure}{0.8\columnwidth}
		\includegraphics[width=\columnwidth]{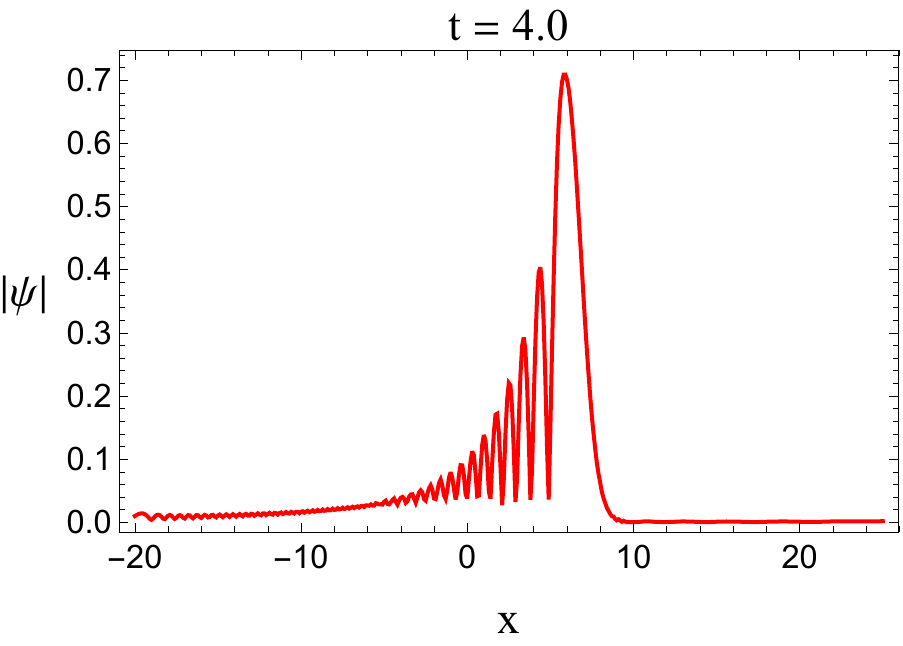}%
		\caption{Solution profile for $ g_1 = -10,g_2=0$ at the time $t=4.0$.}
		\label{t4-g1}%
	\end{subfigure}\hfill%
	\begin{subfigure}{0.8\columnwidth}
		\includegraphics[width=\columnwidth]{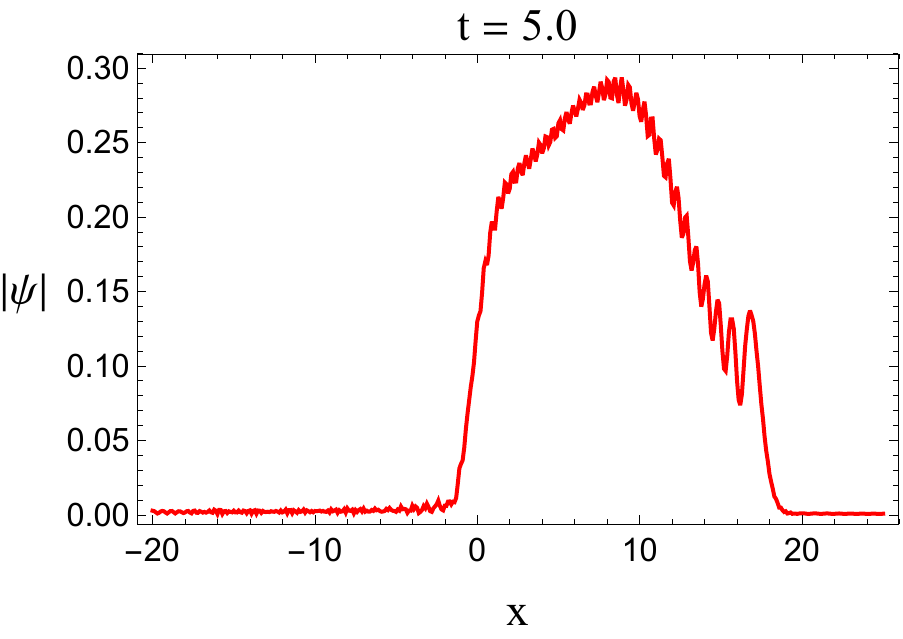}%
		\caption{Solution profile for $ g_1 = -10,g_2=0$ at the time $t=5.0$.}%
		\label{t5-g1}%
	\end{subfigure}\hfill%
	\begin{subfigure}{0.8\columnwidth}
		\includegraphics[width=\columnwidth]{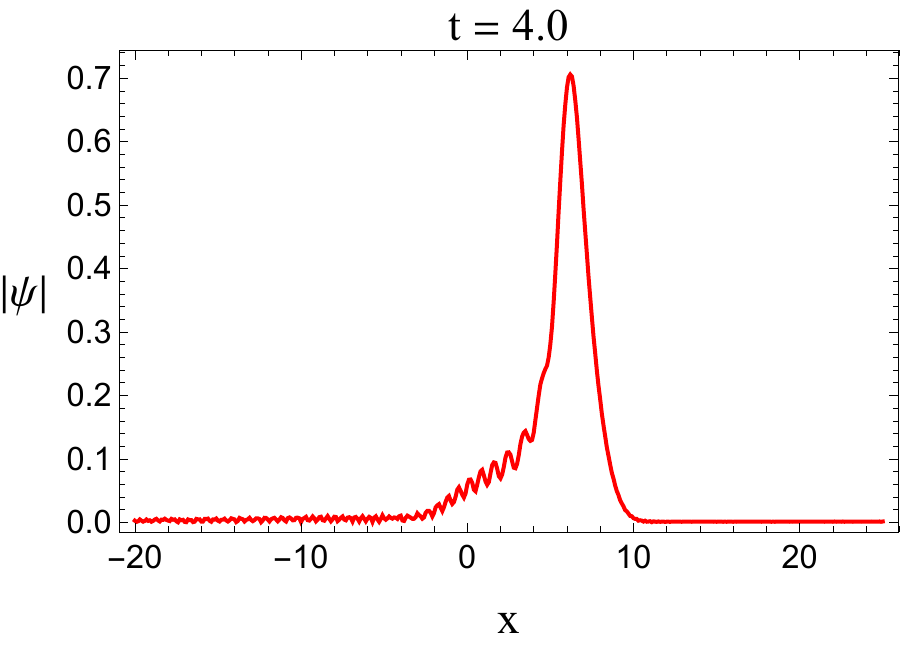}%
		\caption{Solution profile for $ g_1 =0,g_2=10$ at the time $t=4.0$.}%
		\label{t4-g2}%
	\end{subfigure}\hfill
	\begin{subfigure}{0.8\columnwidth}
		\includegraphics[width=\columnwidth]{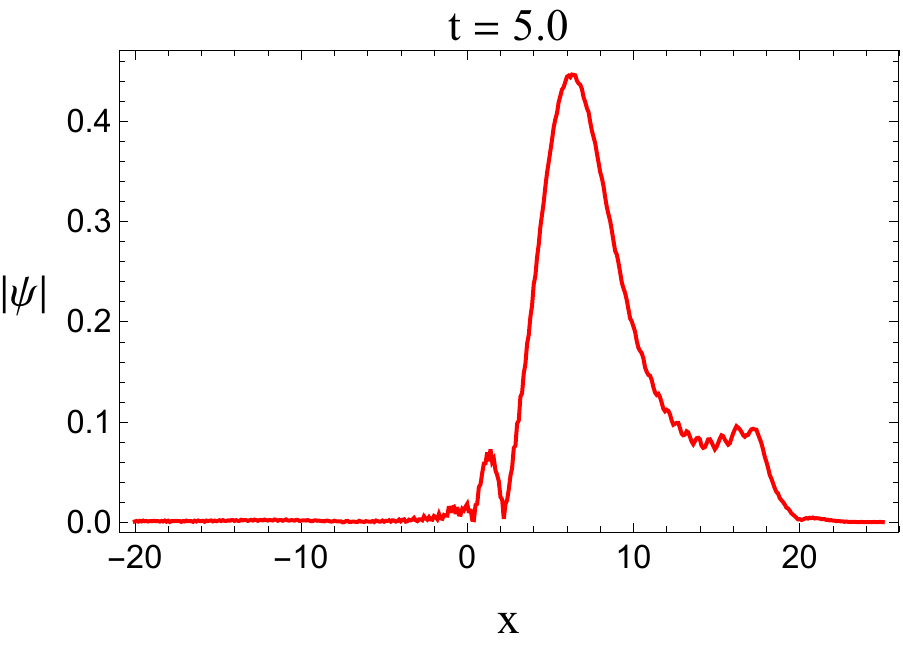}%
		\caption{Solution profile for $ g_1 =0,g_2=10$ at the time $t=5.0$.}%
		\label{t5-g2}%
	\end{subfigure}\hfill%
	\centering
	\caption{ Numerical solutions of the Nonlinear PDE \eqref{1-D-adim-mGPE} for the soliton wave of initial condition \eqref{Gauss-2}. The values used are $A(0) = 1, a(0) = 0.5,c(0) = -0.01, b(0) = 0.1, h(0)= -1, f(0) = -100$ and coefficients $\alpha = 0.00012,\beta = 0.05,\eta= 0.005, K_3= 0.0001, \gamma = 0.00015 ,\omega_0=2\pi/10$. }
	\label{g1-g2}
\end{figure}

Thus, it is interesting to investigate the effect of the higher order nonlinear coefficients $g_1,g_2$ on the movement of the pulse. We note here that Feschbach resonances allow for an increase in the scattering length value, even changing it from attractive $a<0$ to repulsive $a>0$. Now since $g_1=\frac{64\sqrt{2}|a|}{15\pi a_\perp}$ and $g_2=\frac{50a^2}{\pi a^2_\perp}$ we need to take for an attractive interaction a negative $g_1$ but a positive $g_2$.
We start by increasing the $|g_1|$ coefficient with $g_2=0$, so as to isolate the effect of the nonlinear term. The result of such a test is shown in Fig. \ref{t4-g1} and \ref{t5-g1}.The numerical testing shows that increasing the strength of the $|\psi|^3\psi$ term affects the 
%
%
%
%
%
\noindent FR process, dispersing the pulse, increasing the fragmentation, and hindering the regeneration. This can be seen by comparing Fig. \ref{t5-g1} to Fig. \ref{BEC-full-potent} for the same time. Clearly the pulse at $t=5.0$ in Fig. \ref{t5-g1} has more fluctuations at the top, and still some fragmented pulses to the right. The difference is more striking if this is compared to the time independent potential solution. The single-humped behavior of Fig. \ref{BEC-eta0} gets shadowed by the train of pulses. Increasing even further the value of $|g_1|$ only makes the top train of pulses  in Fig. \ref{t5-g1} increase in amplitude and dominate over the form of $|\psi|$, deforming and fragmenting the pulse.

On the contrary, increasing $g_2$ with $g_1=0$ makes the pulse fragmentation less noticeable, as shown in Figs. \ref{t4-g2} and \ref{t5-g2}. The solution for $t=4.0$ differs greatly in comparison to the same time solution for $g_1\neq 0$ in Fig.\ref{t4-g1} or even with small $g_1$, and $g_2$, as in Fig. \ref{BEC-full-potent}. The pulse has little fluctuations on its tails, smaller in amplitude compared to the train of pulses in Fig.\ref{t4-g1}. Moreover, the regeneration of the pulse after the fragmentation is amplified, since the profile \ref{t5-g2} has a greater similarity to the original gaussian function and the non-fragmented solution of Fig. \ref{BEC-eta0} for $t=5.0$. Increasing the value of the quantum fluctuations coming from $g_2$ makes the pulse more robust, and reduces the tendency to disintegrate at least up to $g_2\approx20$. In this way the importance of the nonlinear coefficient $g_2$ manifests as less fragmentation and more regeneration for the pulse. Therefore, the term $|\psi|^4\psi$ in equation \ref{1-D-adim-mGPE} helps the pulse propagation, in comparison to the usual Lee Huang Yang correction of coefficient $g_1$ which tends to destabilize the pulse.

\section{Supergaussian}
\begin{figure}[H]%
	\centering
	\includegraphics[width=0.9\columnwidth]{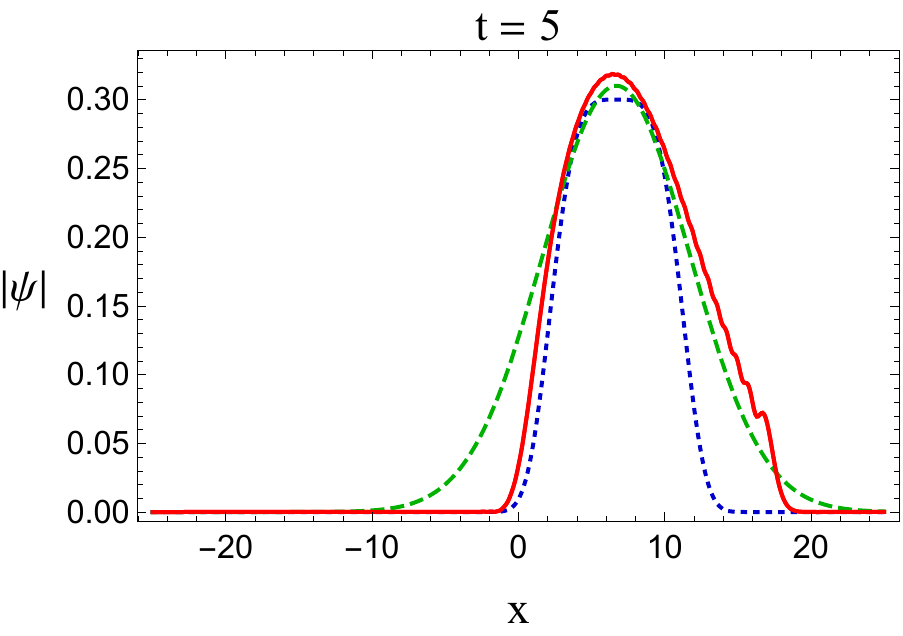}
	\caption{(Color online) Pulse parametrizations, gaussian (green,dashed), supergaussian (blue,dotted) superimposed to the numerical solution of Fig. \ref{BEC-full-potent} at $t=5.0$ in red with a solid line.}
	\label{perfiles}
\end{figure}

In this section we investigate further the connection between the variational approximation and the results from numerical solutions of the PDE \ref{1-D-adim-mGPE}. More precisely: we will investigate if the variational approximation might be improved by using a different trial function, whose shape is closer to pulses such as those shown in Fig. \ref{BEC-full-potent} (at times $t=3$, and $t=5$). In other words, we will use a functional form that goes to zero quicker than a gaussian, and having almost no tails. To this end we introduce a supergaussian trial function of the form:
\begin{align}
\psi(x, t)=A \exp \left\{ -\frac{(x-c)^{4}}{2 a^{2}}+ i\left(b x^{2}+f x+h\right)  \right\},
\end{align}
where $A,c,a,b,f,h$ are functions of the time just as before. A comparison between the different trial functions and the PDE solutions is shown in Fig. \ref{perfiles}.Clearly the part of the pulse with a  steeper inclination is better described by a supergaussian. Next, we introduce  this function on the lagrangian \eqref{lagrangian}, and repeat the variational process. In this case we do not consider the three-body recombination nor the feeding since these constants were seen to be much smaller than all other terms in equation \eqref{1-D-adim-mGPE}, and do not affect its numerical solutions in a significant way. Thus, the Euler Lagrange equations coming from the supergaussian can be put into the form:
\begin{align}
&\dot{c}=4 b c \label{c-super}\\
&\dot{a}=8 b a-\frac{4 \Gamma\left(\frac{1}{4}\right)}{\Gamma\left(\frac{3}{4}\right)} f c \label{a-super}\\
\dot{f}&=2\left(\beta-\dot{b}-4 b^{2}\right) c-4 b f-12 \alpha \frac{\Gamma\left(\frac{3}{4}\right)}{\Gamma\left(\frac{1}{4}\right)} a c-4 \alpha c^{3} \label{f-super}\\
&\dot{b}=\beta-4 b^{2}+\frac{2}{k_{1} \Gamma\left(\frac{3}{4}\right)}\left[\frac{d R}{d a}-\alpha k_{1} \Gamma\left(\frac{5}{4}\right) a-3 \alpha k_{1} \Gamma\left(\frac{3}{4}\right) c^{2}\right.\notag\\
&\left.\hspace{4cm} -\frac{3}{2} k_{1} \eta \operatorname{sen}\left(\omega_{0} t\right) \Gamma\left(\frac{3}{4}\right) c\right] \label{b-super}
\end{align}

\begin{figure}[H]%
	\centering
	\includegraphics[width=0.9\columnwidth]{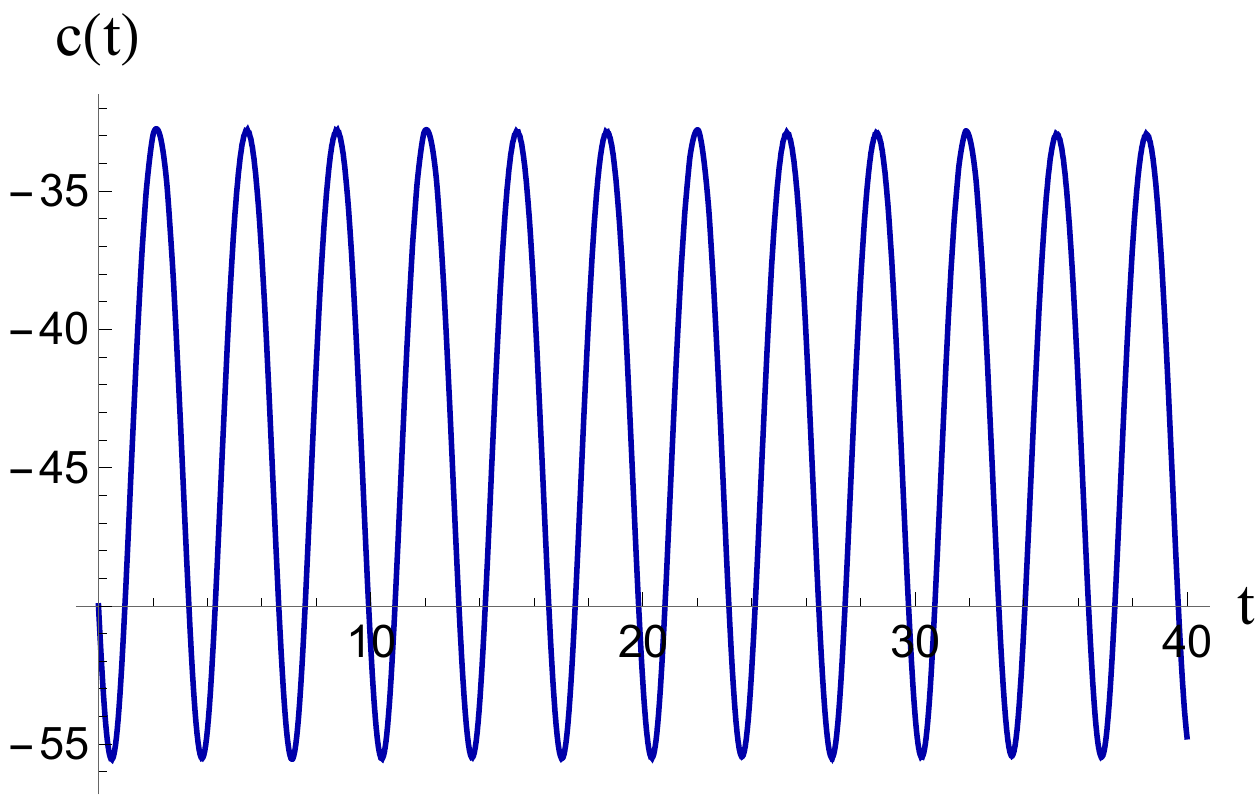}
	\caption{VA numerical solution for the center of mass of a supergaussian trial function with the same initial values as Fig. \ref{VA-same}, and the same initial width at half maximum.}
	\label{supergauss}
\end{figure}

\noindent where the value of $\dot{b}$ from \eqref{b-super} needs to be inserted in eq. \eqref{f-super} and we defined 
\begin{align}
&k_{1}=A^{2} a^{1 / 2} \\
&R(a)=\frac{k_{1}^{2}}{4\left(2^{1 / 4}\right)} \Gamma\left(\frac{1}{4}\right) \frac{1}{a^{1 / 2}}-\frac{k_{1}^{5 / 2} g_{1}}{5\left(\frac{5}{2}\right)^{1 / 4}} \Gamma\left(\frac{1}{4}\right) \frac{1}{a^{3 / 4}} \notag\\
&-\frac{k_{1}^{3} g_{2}}{2\left(3^{5 / 4}\right)} \Gamma\left(\frac{1}{4}\right) \frac{1}{a}-2 k_{1} \Gamma\left(\frac{7}{4}\right) \frac{1}{a}
\end{align}

As before, we plot in Fig.\ref{supergauss} the center of mass trajectory as a function of time. The parameters were chosen so as to have the same width at half maximum as the Gaussian expression, with the same amplitude and initial values in the phase. By comparing the V.A. prediction with the solution of the PDE shown in Fig. \ref{PDE-same} it is seen that the fine bumps of the pulse are not described by the supergaussian. Consequently, it is seen that the Gaussian profile is a better trial function than a supergaussian, at least for describing the overall movement of the center of mass of a BEC's pulse which obeys Eq. \ref{1-D-adim-mGPE}.

\section{Conclusions}

In this paper we studied the dynamical properties of soliton-like solutions for a high-density condensate in the presence of quantum fluctuations, atomic feeding and three body losses. The quantum fluctuation we investigated had a novel extra term coming from the expansion of the energy in high density states. This condensate is confined in a time oscillating potential that could be realized with electromagnetic or optical systems. By employing a variational approximation, we deduced a set of coupled ODEs describing the evolution of a trial gaussian function. The center of the soliton presented an oscillating behavior which involves  two frequencies: a high frequency which is determined by the fourth order term of the potential, and the fre-

\noindent quency of the envelope of the rapid oscillations, which coincides with the frequency of the third-order term of the potential. 

The initial conditions of the pulse such as the  center position, the velocity and the width seemed to change dramatically the solution, while the position independent phase and chirp were extraneous to the BEC behavior over the same timescales. For a center position near a minimum of the potential the variational approximation showed an excellent agreement with the BEC numerical solution, specially predicting the location of extrema in the amplitude and center of mass. Two trial functions were used: a gaussian, and a supergaussian. While the latter matched the profile of the numerical solution  better, it was found that only gaussian function followed the center of mass trajectory. The exemplary agreement between the Anderson-Hasegawa variational method, and the solutions near a minimum of the potential verified the validity of the numerical calculation and the usefulness of the VA, for a single hump pulse.

However, when the condensate passes over the potential barrier at $x=0$ and oscillates between the two wells, the VA is able to describe the gross characteristics of the BEC's behavior, but the direct numerical solutions of Eq. (4) shows that the BEC's shape presents a fine structure, which in Sec. 4 was described as a “fragmentation-regeneration (FR) process”. The occurrence of this FR process is an interesting result, as it cannot be predicted by the VA, nor we could have imagined it in advance. In addition, the FR process survives a time varying perturbation as long as the variation presents itself in the anharmonic term. This behavior seems to depend strongly on the coefficients of the potential and the magnitude of the quantum fluctuations. An increase in the fourth-order coefficient of the potential, and/or an increase in the third-order coefficient, led to the destruction of the pulse. On the contrary, an increase in the novel nonlinear term $g_2$ helps the regeneration and reduces fragmentation of the pulse, in contrast with the effect of increasing the value of $|g_1|$, which is detrimental for the FR process.
\section*{Acknowledgments}

We thank DGTIC-UNAM (Dirección General de Cómputo y de Tecnologías de Información y Comunicación de la Universidad Nacional Autónoma de México) for granting us access to the supercomputer \textit{Miztli} through Project LANCAD-UNAM-DGTIC-164, in order to carry out this work.

\interlinepenalty=10000
\bibliography{FR-BEC-RV-bibliog}
	
\end{document}